\documentclass[english,aps,twocolumn,showpacs,showkeys,preprintnumbers,prd,floatfix,nofootinbib,superscriptaddress,notitlepage,hidelinks]{revtex4-2}

\usepackage{graphicx}
\usepackage{amsmath}
\usepackage{amsfonts}
\usepackage{hyperref}
\hypersetup{
    colorlinks = true,
    linkcolor = blue,
    urlcolor  = blue,
    citecolor = blue,
    anchorcolor = blue,
    pdftitle = {Determination of proton PDF uncertainties with Markov Chain Monte Carlo},
    pdfauthor={Risse, P. et al.}
    }
\usepackage{color}
\usepackage{placeins}
\usepackage{tabu}
\usepackage{nicefrac}
\usepackage{ulem}
\usepackage{float}
\usepackage[caption=false]{subfig}
\usepackage[capitalise]{cleveref}
\usepackage{bm}
\usepackage{duckuments}

\makeatletter
\DeclareRobustCommand\em
  {\@nomath\em \ifdim \fontdimen\@ne\font >\z@
     \eminnershape \else \itshape \fi}%
\makeatother

\newcommand{\orcid}[1]{\,\href{https://orcid.org/#1}{\includegraphics[width=9pt]{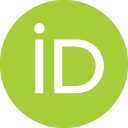}}\,}

\newcommand{\MCMCparam}{\ensuremath{\bm{X}}}
\newcommand{\x}{\ensuremath{\bm{X}}}
\newcommand{\mub}{\ensuremath{\bm{\mu}}}
\newcommand{\Ob}{\ensuremath{\mathcal{O}}}

\newcommand{\asym}{$\alpha$\%-symmetric}
\newcommand{\aasym}{$\alpha$\%-asymmetric}
\newcommand{\cumchi}{cumulative-$\chi^2$}

\newcommand{\orcidPR}{0000-0002-8570-5506} 
\newcommand{\orcidTJ}{0000-0002-1334-7607} 
\newcommand{\orcidKK}{0000-0003-1412-447X} 
\newcommand{\orcidAK}{0000-0002-4090-0084} 
\newcommand{\orcidND}{0000-0003-0962-631X} 

\newcommand{\muenster}{\affiliation{Institut f{\"u}r Theoretische Physik, Universit{\"a}t M{\"u}nster,
Wilhelm-Klemm-Stra{\ss}e 9, D-48149 M{\"u}nster, Germany}}
\newcommand{\smu}{\affiliation{Department of Physics, Southern Methodist University, Dallas, TX 75275-0175, U.S.A.}}
\newcommand{\jlab}{\affiliation{Theory Center, Jefferson Lab, Newport News, VA 23606, U.S.A.}}

\newcommand{\krakow}{\affiliation{Institute of Nuclear Physics Polish Academy of Sciences, PL-31342 Krakow, Poland}}

\begin{document}
\preprint{
\vbox{
\null \vspace{0.3in}
\hbox{SMU-PHY-25-02, JLAB-THY-25-4575, MS-TP-25-29, IFJPAN-IV-2025-17}
}}

\title{
Determination of proton PDF uncertainties with Markov Chain Monte Carlo}

\author{P.\,Risse\orcid{\orcidPR}}
\email{prisse@smu.edu}
\muenster\smu\jlab

\author{N.\,Derakhshanian\orcid{\orcidND}}
\krakow 

\author{T.\,Je\v{z}o\orcid{\orcidTJ}}
\email{tomas.jezo@uni-muenster.de }
\muenster

\author{K.\,Kova\v{r}\'{\i}k\orcid{\orcidKK}}
\muenster

\author{A.\,Kusina\orcid{\orcidAK}}
\email{aleksander.kusina@ifj.edu.pl}
\krakow 

\date{\today}

\begin{abstract}
\vspace*{0.5cm}
We present an analysis of parton distribution functions (PDFs) of the proton using Markov Chain Monte Carlo (MCMC) methods. The MCMC approach naturally implements Bayes' theorem and thus provides a means to directly sample the underlying probability distribution -- in this case the probability distribution of the PDF parameters. This allows for a straightforward propagation of the resulting uncertainties into any PDF-dependent observable, preserving their simple probabilistic interpretation. 
In our analysis we include a broad set of deep inelastic scattering data from HERA, BCDMS and NMC experiments along with the Drell-Yan, $W$ and $Z$ boson data from LHC and Tevatron experiments, which combined with theoretical calculations at next-to-next-to-leading order in QCD allow for realistic determination of PDFs.
The main focus of this analysis is to explore alternative methods for PDF uncertainty estimation that are more firmly grounded in statistical principles. We show that the flexibility of the Bayes framework, allowing e.g.~to account for non-Gaussianity or inconsistencies of data sets, is crucial to extract realistic uncertainties when such assumptions are not fulfilled. We also demonstrate that MCMC allows one to determine the $\Delta \chi^2$ value corresponding to a given confidence level in the sample, which can in turn be used as a statistically well-founded tolerance criterion used in the Hessian method, thus addressing one of its main long-standing drawbacks. 
\end{abstract}

\maketitle
\tableofcontents{}

\section{Introduction}
\label{sec:intro}
We know that the Standard Model of particle physics (SM) is not a complete theory, yet the available collider data agrees remarkably well with its predictions. This is particularly apparent looking at the wealth of the Large Hadron Collider (LHC) data spanning many orders of magnitude in cross sections~\cite{ATL-PHYS-PUB-2024-011}. 
As a result, the most promising opportunity for searches of new physics in the coming high-luminosity LHC runs is through high precision comparisons of theoretical predictions with experimental measurements. For this purpose the uncertainties of both the data and theory predictions need to be comparable, and already now in many cases the theory uncertainties dominate. For this reason, we need to scrutinize the sources of theory uncertainties, take steps to reduce them and make their estimates as reliable as possible. There are several key sources of theory uncertainties, with two of the most significant being: (i)~related to the truncation of the perturbative expansion commonly referred to as missing higher-orders (MHOs) uncertainties, (ii)~related to parton distribution functions (PDFs).
In addition to these, non-perturbative effects such as hadronization, the creation of hadron jets, and the formation of resonances contribute to uncertainties.
The uncertainties due to MHOs are typically estimated using scale variations; recently also other methods were proposed, e.g., theory nuisance parameters~\cite{Tackmann:2024kci,Lim:2024nsk,Cacciari:2011ze,Charles:2016qtt,Bagnaschi:2014wea,Bonvini:2020xeo,Duhr:2021mfd}. In this work, we will concentrate on PDFs and their uncertainties.

The PDF related uncertainties are particularly important in view of the fact that they make up the dominant part in the theory error budget.
It is widely understood that these uncertainties need to be reduced and reliably estimated~\cite{NNPDF:2021njg,Gao:2017yyd,Ubiali:2024pyg} in order to meaningfully enhance our confidence in the SM or advance the searches of new physics in the coming years. Since PDFs are non-perturbative quantities the first-principle calculations of PDFs are very complicated or even impossible across the whole range of momentum fractions of interest~\cite{Lin:2017snn,Cichy:2018mum,Constantinou:2020hdm} and in practice, for phenomenological applications, they are determined by fitting experimental data~\cite{Gao:2017yyd,Kovarik:2019xvh,Ethier:2020way,Alekhin:2017kpj,NNPDF:2021njg,Hou:2019efy,Bailey:2020ooq,McGowan:2022nag,Cerutti:2025yji,H1:2015ubc}. As a result the main source of uncertainties in phenomenologically determined PDFs originates from the experimental data used in the fits.%
\footnote{Of course, theoretical uncertainties, such as MHOs, as well as methodological uncertainties also contribute.}

In this paper we will not try to directly reduce the current PDF uncertainties but in turn focus on their reliable estimation. For this purpose we propose to use Markov Chain Monte Carlo (MCMC) methods to determine the PDFs. We will show that this method allows for a robust estimation of PDF uncertainties and their propagation into PDF-dependent observables, allowing for a simple and well-defined probabilistic interpretation. We will also contrast the error estimates obtained from MCMC with those obtained with one of the commonly used methods, the Hessian method.

We note that the use of MCMC methods in the context of PDF determination has been attempted before. 
In 2001, Giele et al.~\cite{Giele:2001mr} carried out the first determination of proton PDFs using MCMC methods, notable also because they used MCMC to infer detector effects as well. Later, in Ref.~\cite{Gbedo:2017eyp}, the first modern toy study using only HERA data was conducted, focusing on the up- and down-valence distributions. 
Building on these early efforts, the authors of Refs.~\cite{Hunt-Smith:2022ugn, Hunt-Smith:2023ccp}, performed a comparison of various uncertainty estimation methods, including the Hessian method, data resampling, MCMC, and nested sampling, and introduced improvements to the MCMC algorithm for better sampling efficiency, which they demonstrated using a PDF toy model example.
More recently, a MCMC package tailored to PDF determination was introduced~\cite{Capel:2024qkm}, using a forward model approach where the number of counts per bin is predicted directly rather than relying on unfolded cross sections. Their analysis, which is limited to deep inelastic scattering processes, reports that the evaluation of physical processes remains too slow for a full-scale analysis.
Finally, a new tool for PDF determinations, called \texttt{Colibri}~\cite{Costantini:2025agd}, has recently been published with an emphasis on comparing different methods in a consistent setup for PDF uncertainty estimations. One of the available methods allows for a Bayesian inference implemented through a nested sampling algorithm. It has been developed in parallel with a novel approach for simplifying PDF models through an orthogonal decomposition of the functional space~\cite{Costantini:2025wxp}. This procedure yields a PDF model linear in its parameters, which simplifies the Bayesian inference, as shown in a fit to global DIS data in a closure test framework.

This paper is organised as follows. Standard approaches to PDF uncertainty estimation, including the Hessian method, the Monte Carlo replica method and other alternatives, are briefly reviewed in \cref{sec:unc}. Section~\ref{sec:setup} describes the analysis setup, including the selected datasets, theoretical predictions and the input PDF parametrization. Results obtained using MCMC methods are presented in \cref{sec:MCMC_vs_Hess}, with a focus on uncertainty estimates and their comparison to conventional approaches. Finally, \cref{sec:conclusions} discusses the main findings and outlines potential methodological improvements and future directions.

\section{Uncertainties in PDFs}
\label{sec:unc}

In this section we review currently available methods for estimating PDF uncertainties. We focus here on the propagation of experimental uncertainties into the PDFs, and will not discuss approaches incorporating other sources of uncertainties.%
\footnote{Propagation of theoretical uncertainties is mostly focused on uncertainties related to the strong coupling or heavy-quark masses, e.g.~\cite{Alekhin:2017kpj,Forte:2020pyp,Gao:2013wwa,Ball:2025xgq,Ablat:2025gbp,Cridge:2024exf}. More recently uncertainties due to missing higher orders were also investigated~\cite{NNPDF:2019vjt,NNPDF:2019ubu,McGowan:2022nag,Kassabov:2022orn,NNPDF:2024dpb}. Some PDF groups also try to estimate ``model'' uncertainties, e.g., related to parametrization~\cite{Alekhin:2014irh,HERAFitterdevelopersTeam:2014fzy}.}

\subsection{Hessian method}
\label{subsec:Hess}
One of the widely used approaches to propagate the experimental uncertainties into PDF analyses is the Hessian method~\cite{Pumplin:2000vx,Pumplin:2001ct,Martin:2002aw}. This method has been used in the context of PDF fits for over 20 years by many different groups working on global analyses~\cite{Pumplin:2002vw,Hou:2019efy,Martin:2002aw,Bailey:2020ooq,H1:2015ubc,Alekhin:2013nda,Gluck:2007ck,Accardi:2016qay}. The biggest advantage of this approach is its relative simplicity. At the same time it faces certain limitations that become more important as the demands on the precision of PDFs grow.

A typical PDF analysis relying on the Hessian method begins by defining an appropriate $\chi^2$ function that compares theoretical, PDF-dependent, predictions with the measured data. The PDFs are parametrized at a low initial scale, $\mu_0$, and the $\chi^2$ function is minimized with respect to the PDF parameters, here denoted as $\{p_i\}$. The values of these parameters at the minimum, $\{p_i^0\}$, define the {\em central} PDFs. 
Next, the experimental uncertainties are propagated.
The $\chi^2$ function is expanded around the minimum, up to the second order:
\begin{equation}
    \chi^2 = \chi^2_{\mathrm{min}}
    + \sum_{i,j} \frac{1}{2} \frac{\partial^2\chi^2}{\partial p_i \partial p_j}\bigg|_{\mathrm{min}} (p_i-p_i^0)(p_j-p_j^0),
\end{equation}
where the matrix of second derivatives is referred to as the Hessian matrix, $H_{ij}=\frac{1}{2} \frac{\partial^2\chi^2}{\partial p_i \partial p_j}\big|_{\mathrm{min}}$. The inverse of the Hessian is the error matrix which after diagonalization is used to explore the vicinity of the minimum.
Uncertainties on fitted parameters are then defined by the region in parameter space where the increase in the $\chi^2$ function stays below a certain threshold. Specifically, the uncertainty region corresponds to the contour $\Delta\chi^2 = T$, where $T$ is determined from Wilks' theorem~\cite{Wilks:1938dza} based on the number of fitted parameters. This construction assumes that the experimental uncertainties are Gaussian and that the data are mutually consistent with the theoretical model. Under these conditions, $T$ is uniquely defined by the desired confidence level and the dimensionality of the parameter space.

However, in practice, these ideal assumptions are rarely fully satisfied. Deviations from Gaussianity, tensions between datasets, and the presence of theoretical uncertainties all lead to a breakdown of the strict statistical interpretation. As a result, the choice of the $\Delta\chi^2$ tolerance becomes somewhat arbitrary and is often adjusted in an ad hoc manner to account for these imperfections. Probably the biggest limitation of the Hessian method is this very choice: the tolerance defines the vicinity of the minimum that is explored in order to quantify the uncertainties, but in practice there is no unique or theoretically well-motivated way to set it. Instead, the appropriate tolerance must be chosen depending on the specific dataset and the fit quality.

Since its introduction~\cite{Pumplin:2001ct}, the Hessian method has evolved, and modern PDFs relying on it~\cite{Hou:2019efy,Bailey:2020ooq,Eskola:2021nhw,Duwentaster:2022kpv} employ many refinements, such as dynamic tolerance criterion~\cite{Martin:2009iq}.
Nevertheless, the main ``problems'' -- the choice of the tolerance criterion and assumptions about Gaussian errors -- are unavoidable and can limit the precision this method can deliver.

\subsection{Monte Carlo replica method}
\label{subsec:MCrep}

In the beginning of 2000s an alternative to the Hessian method was proposed -- the Monte Carlo replica method~\cite{Forte:2002us,DelDebbio:2007ee} -- which we will refer to as the Monte Carlo or the replica method. Instead of finding the {\em central} PDF corresponding to a global minimum of a certain loss/$\chi^2$ function, and exploring the vicinity of this minimum to estimate the uncertainties, the key idea of this method is to find a set of equally probable PDFs. To this end replicas of experimental data are created~\cite{DelDebbio:2007ee}, allowing for fluctuations consistent with experimental uncertainties. Then, for each of those replicas, a minimum in PDF parameter space is found providing a set of PDF replicas that can be used to estimate PDF uncertainties of any PDF-dependent observable.
The NNPDF group, which introduced the replica method, also combines it with the use of very flexible parametrizations through the use of neural networks (NN)~\cite{NNPDF:2021njg,NNPDF:2017mvq} which further distinguishes it from the Hessian method which is typically limited to fixed functional parametrization forms.
The predictions and uncertainties for any observable (including PDFs themselves) can be computed as mean and variance taken on PDF replicas, or alternatively by constructing confidence intervals~\cite{AbdulKhalek:2022fyi}.

Recent considerations of the Monte Carlo method ~\cite{Costantini:2024wby} find that the posterior obtained with the replica method agrees with the posterior from more general Bayesian approaches only for linear models (and a sufficiently wide Bayesian prior). In general, linearity in PDF fits is broken in several aspects. The most obvious example is the inclusion of experimental data from hadron colliders where the theory prediction consists of a product of PDFs regardless of the parametrization.\footnote{One could argue that linearity only matters in the vicinity of the best fit estimate, but we find below that at least some of the PDF parameters in our setup are not distributed in a linear fashion, not even in the close vicinity of the minimum.} The authors of the reference above show that the Monte Carlo replica method introduces distortions of the posterior that are very difficult to predict, even in toy models, and can have a significant impact. Nevertheless, the extent of this discrepancy in realistic PDF fits remains, to our knowledge, an open question.
Moreover, the use of a neutral network as part of the PDF parametrization implies the split of experimental data into a training and a validation set. For linear models it can be shown that such a spilt leads to inflated uncertainties on the extracted quantities, which is easily understood because the split removes data constraints. It is possible to demonstrate this effect on non-linear models, but it remains poorly understood, due to the complicated distortions of the posterior introduced by the replica method~\cite{Costantini:2024wby,Hunt-Smith:2022ugn}.

\subsection{Other methods}
\label{subsec:Other}

Several alternative approaches for propagating experimental uncertainties into PDFs have been proposed to address the limitations of the Hessian and the replica methods. 
One such approach is the Lagrange multiplier method~\cite{Stump:2001gu,Martin:2002aw}, which investigates how the global $\chi^2$ changes as specific observables are varied, allowing a direct assessment of the impact of individual measurements on the fit. 
Unlike the Hessian method, it does not rely on a quadratic approximation of the $\chi^2$ near the minimum.

More recently, neural network–based techniques have been developed to model the structure of the likelihood function more flexibly. 
For example, a method introduced in Ref.~\cite{Liu:2022plj} trains a NN on an effective posterior distribution, making it possible to represent non-Gaussian uncertainties and complex parameter correlations without committing to specific functional forms. 

Another promising direction involves the use of Gaussian processes to represent the dependence of observables on PDFs. 
As shown in Ref.~\cite{Candido:2024hjt}, this non-parametric Bayesian approach provides a probabilistic interpolation between theoretical predictions, naturally incorporating uncertainty estimates and correlations in a flexible framework. 

Finally, MCMC methods provide a fully Bayesian treatment of uncertainties by sampling directly from the posterior distribution of fit parameters. 
Although not widely adopted in global PDF analyses, several studies, such as~\cite{Gbedo:2017eyp}, have successfully applied them to extract detailed uncertainty estimates and reveal non-linear features that are inaccessible in Hessian-based frameworks. 
This is the methodology we adopt in the present study, as it enables a statistically grounded exploration of the full posterior and avoids reliance on ad hoc tolerance criteria or Gaussian approximations.

\section{Analysis setup}
\label{sec:setup}

This section describes the setup of our analysis starting from the selection of experimental data sets, the treatment of theoretical predictions and the PDF parametrization. Finally, we describe the MCMC analysis providing both an introduction to the method and listing setups specific to our fit. Note that in this exploratory study we limit the sophistication of our setup in order to keep the computational costs manageable. Thus several compromises have been made, e.g.~the selection of experimental data sets or the choice of MCMC algorithm. Our goal is to benchmark the uncertainty estimate obtained using MCMC against traditional alternatives.

\subsection{Data selection}
\label{subsec:data}

Without loss of generality, we restrict our data selection to two processes: deep inelastic scattering (DIS) and Drell-Yan lepton pair production (DY). The DIS data includes the combined H1 and ZEUS data from the HERA experiment (in the form of reduced cross section for neutral and charged current) and $F_2$ structure function data from the BCDMS and NMC experiments. The DY data includes $Z$ and $W^{\pm}$ boson production data from Tevatron and LHC. 
To minimize the effects of higher twist contributions we apply kinematic selection cuts. In case of the DIS data we use: $Q\geq 2$ GeV and $W\geq 3.5$ GeV, and for DY data we follow the choices made in~\cite{NNPDF:2017mvq}.%
\footnote{Since we also use fast convolution grids provided in~\cite{NNPDF:2017mvq} it is natural to follow the same cuts. 
The cuts are ATLAS $Z$ $p_T$ 8 TeV ($y_Z$): $30\,{\rm GeV}\leq p_T\leq 150\,{\rm GeV}$, CMS $Z$ $p_T$ 8 TeV: $20\,{\rm GeV}\leq p_T\leq 170\,{\rm GeV}$, CMS double diff. 2011 7 TeV $y\leq2.2$ and LHCb $W^\pm,Z\rightarrow \mu$ 7 \& 8 TeV: $2.25\leq \eta$.}
The total number of data points after (before) kinematic cuts is equal to 1984 (2427), with the majority [1660 (1949)] comprised of the DIS data. The remaining 324 (478) data points are for the DY process with most of them originating from the LHC. Table~\ref{tab:dataSets} summarizes individual data sets together with the number of points.
\begin{table}[ht]
    \centering
    \begin{tabular}{lcc}
    \textsc{Data Set}  & \textsc{Ref.} & \textsc{Data Points} \\
    \hline
    \textbf{DIS} & & \\
    HERA $\sigma_{red}$ neutral current & \cite{H1:2015ubc} & 1039 (1225)\\
    HERA $\sigma_{red}$ charged current & \cite{H1:2015ubc} & 81 (81)\\
    BCDMS $F_2$ proton & \cite{BCDMS:1989qop} & 339 (351)\\
    NMC $F_2$ proton & \cite{NewMuon:1996fwh} & 201 (292)\\[4pt]
    DIS total & & 1660 (1949)\\[4pt]
    \textbf{DY} & & \\
    CDF $Z$-rapidity & \cite{CDF:2010vek} & 28 (28) \\
    D{\O} $Z$-rapidity & \cite{D0:2007djv} & 28 (28)\\
    ATLAS $Z$ $p_T$ 8 TeV ($M_{ll}$) & \cite{ATLAS:2015iiu} & 44 (64)\\
    ATLAS $Z$ $p_T$ 8 TeV ($y_Z$) & \cite{ATLAS:2015iiu} & \,\,48 (120)\\
    CMS $Z$ $p_T$ 8 TeV & \cite{CMS:2015hyl} & 28 (50)\\
    CMS double diff. 2011 7 TeV & \cite{CMS:2013zfg} & \,\,88 (120)\\
    LHCb $W^\pm,Z\rightarrow \mu$ 7 TeV & \cite{LHCb:2015okr} & 29 (33)\\
    LHCb $W^\pm,Z\rightarrow \mu$ 8 TeV & \cite{LHCb:2015mad} & 31 (34)\\[4pt]
    DY total & & 324 (478)\\[4pt]
    \hline
    \textbf{Total} & & 1984 (2427)
\end{tabular}
    \caption{The experimental data sets along with the number of data points used for the PDF extraction. The number in brackets is the total number of data points before applying kinematical cuts.}
    \label{tab:dataSets}
\end{table}

\begin{figure*}[ht!]
    \centering
    \includegraphics[width=0.99\textwidth]{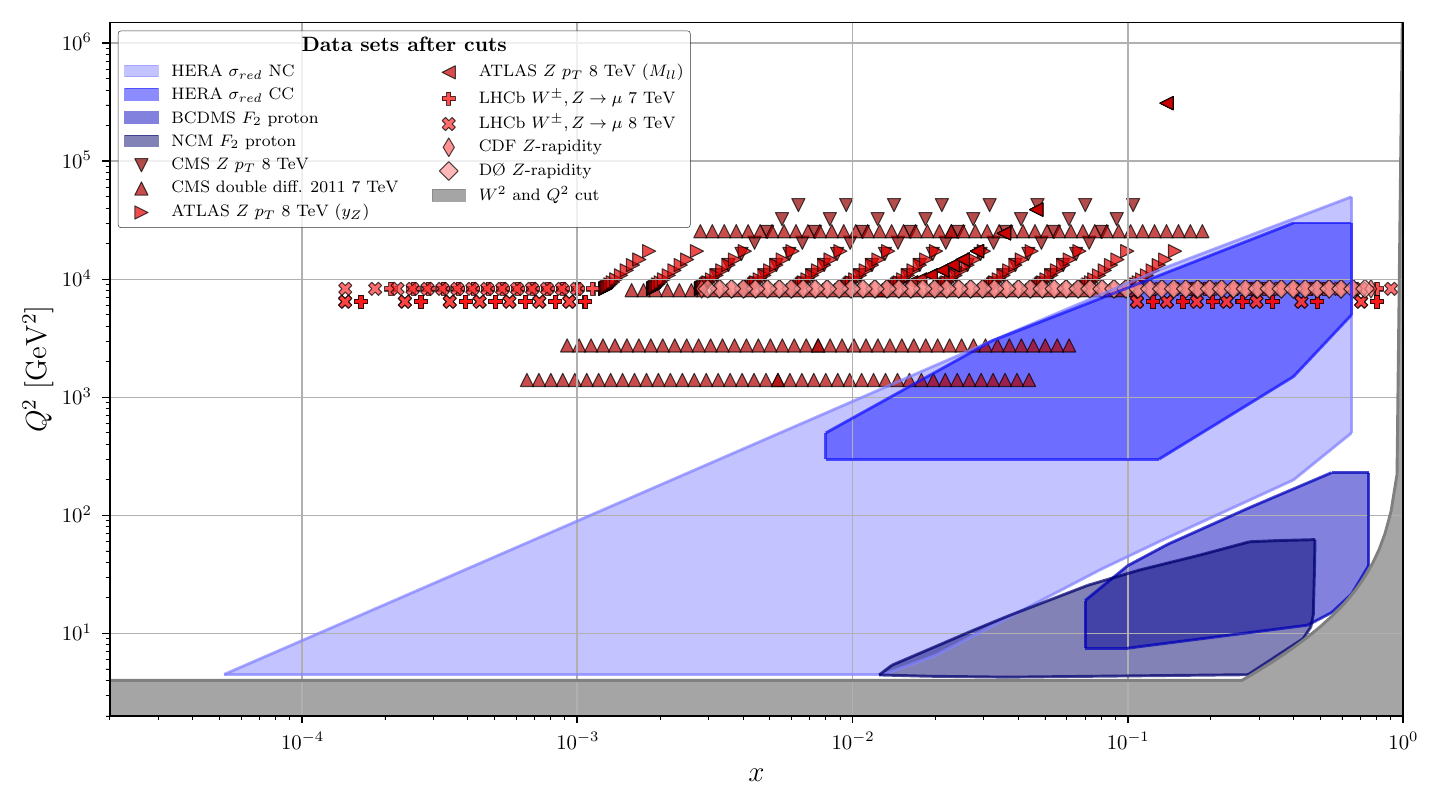}
    \caption{Kinematic coverage of the experimental data sets in the $(x,Q^2)$-plane. The data sets are displayed after cuts and leading order kinematics have been assumed for the DY data sets. The gray area indicates the global $Q^2$- and $W^2$-cut.}
    \label{fig:kinematic_coverage}
\end{figure*}
Additionally, in Fig.~\ref{fig:kinematic_coverage} we show the kinematic coverage of the fitted data in the $(x,Q^2)$ plane. We can see that the HERA data provides a wide coverage in $x$ from very low $x\sim5\times10^{-5}$ up to large $x\sim0.5$. The fixed target DIS data provide additional constraints in the large $x$ region. The $Z$ and $W^{\pm}$ boson production data also cover a relatively wide range in $x$, going down nearly to $10^{-4}$ and extending nearly to $x\sim1$ but at a higher $Q^2$ on average.
A selection such as this is expected to provide reliable constraints on the valence and light-sea ($\bar{d}+\bar{u}$) quark distributions. 
The limited amount of flavour changing observables restricts the separation of the sea-quarks reflected in the largely fixed parameters of the $\bar{d}/\bar{u}$ and $s+\bar{s}$ combinations.
However, even though none of the used data are directly sensitive to the gluon distribution at LO they do allow to constrain the gluon PDF via higher order effects. The large lever arm of the HERA data provides constraints coming from the DGLAP evolution and the high-energy used at the LHC enhances the sensitivity of the DY data.

\subsection{Theory predictions}
\label{subsec:thory}

Theoretical predictions for neutral current (NC) DIS data is performed using the aSACOT-$\chi$ scheme~\cite{Stavreva:2012bs,Risse:2025smp} at approximate NNLO (aNNLO). The aSACOT-$\chi$ scheme is an approximate extension of the general mass variable number ACOT scheme~\cite{Aivazis:1993kh,Aivazis:1993pi,Kretzer:1998ju,Kramer:2000hn} to NNLO and allows (with simplifying assumptions) to account for heavy-quark mass effects. With the kinematics of the employed DIS data the mass effects are important only for the NC data. In the case of the charged current (CC) DIS data, which comes only from HERA, the difference between using a massive or massless scheme is negligible ($\sim 0.2$ points in $\chi^2$ for the 81 data points). Hence we use the zero-mass variable-number scheme (ZM-VFNS)~\cite{Collins:1986mp} for the CC DIS data as the evaluation of the massless predictions is computationally less expensive. In both NC and CC case we use tabulated predictions in form of grids provided in the \texttt{APFEL++} code~\cite{Bertone:2017gds}. The \texttt{HOPPET} code is used to carry out the DGLAP evolution~\cite{Salam:2008qg}.

Theoretical calculations for the DY process are time consuming already at NLO as a result they need to be sped up to be used in any PDF analysis. This is even more relevant for the current MCMC analysis. For this reason the calculations are done using the MCFM program~\cite{Campbell:2011bn} interfaced with APPLgrid~\cite{Carli:2010rw} allowing to produce interpolating tables/grids with NLO predictions.%
\footnote{We use APPLgrids that were published by the NNPDF collaboration \href{https://github.com/NNPDF/applgrids}{https://github.com/NNPDF/applgrids}.}
To obtain even further speed up the APPLgrids are then converted to FK-tables~\cite{Bertone:2016lga} which also include the DGLAP evolution from the \texttt{APFEL}-code~\cite{Bertone:2013vaa,Carrazza:2014gfa} into the grids. To obtain the NNLO predictions the K-factors published in~\cite{NNPDF:2017mvq,nnpdfKfac} are used.
With this setup for the DIS and DY theory calculations the computation time for a single $\chi^2$-evaluation for all 1984 data points is $\sim 0.4$ seconds (on the PALMA II cluster~\cite{palma2}).

\subsection{Definition of likelihood and $\chi^2$ function}
\label{subsec:chi2}

We use the standard definition of the $\chi^2$ function, see e.g.\,Ref.\,\cite{Kovarik:2019xvh}, which accounts for correlated uncertainties within data sets:
\begin{equation}
	\chi^2(D,T(\{p_k\})) = \sum_{ij}^{N_D}\big[D_i-T_i(\{p_k\})\big]C^{-1}_{ij}\big[D_j-T_j(\{p_k\})\big]
\end{equation}
where $D_i$ are the measured values, $T_i$ the corresponding theory calculations depending on the PDF parameters $\{p_k\}$, $C_{ij}$ is the experimental covariance matrix, and $N_D$ is the total number of data points.
Additionally, to avoid d’Agostini bias~\cite{DAGOSTINI1994306} the normalization uncertainties of the data are treated following~\cite{Muzakka:2022wey} (see App.~B) which is equivalent to the $t$-method~\cite{Ball:2009qv}.%
\footnote{The above normalization treatment is applied to the following data sets: BCDMS $F_2$ proton, NMC $F_2$ proton, CDF $Z$-rapidity, ATLAS $Z$ $p_T$ 8 TeV ($M_{ll}$), ATLAS $Z$ $p_T$ 8 TeV ($y_Z$), CMS $Z$ $p_T$ 8 TeV, CMS double diff. 2011 7~TeV, LHCb $W^\pm,Z\rightarrow \mu$ 7 TeV, LHCb $W^\pm,Z\rightarrow \mu$ 8 TeV.}
This $\chi^2$ function is then used to define the likelihood function that will be used (multiplied with our choice of priors, \textit{cf. }\cref{subsubsec:prior}) in the MCMC algorithm:
\begin{equation}
	l(\{p_k\}\,|D) \propto \exp\left(-\frac{1}{2}\chi^2\right)\,.
    \label{eq:likelihood}
\end{equation}
Note, that the proportionality constant is irrelevant for the parameter-optimization with MCMC methods used here.

\subsection{PDF parametrization}
\label{subsec:param}

We parametrize PDFs at the initial scale, $Q_0$, set equal to the charm quark mass, $Q_0=m_c=1.3$ GeV. We use the following functional form inspired by the CJ15 PDF analysis~\cite{Accardi:2016qay}: 
\begin{equation}
    xf_i(x,Q_0) = p_0^{i}x^{p_1^{i}}(1-x)^{p_2^{i}}\left(1+p_3^{i}\sqrt{x}+p_4^{i}x\right)
\label{eq:parm}
\end{equation}
for $i=u_v,\bar{d}+\bar{u},g,s+\bar{s}$.
For $d_v$ the functional form also depends on the $u_v$ distribution:
\begin{equation}
\begin{split}
    xd_v(x,Q_0) &= p_0 \big[ x^{p_1}(1-x)^{p_2} \left(1+p_3\sqrt{x}+p_4x\right)
    \\&
    + p_5x^{p_6} \, xu_v(x,Q_0)\big]\,.
\label{eq:parmdv}
\end{split}
\end{equation}
The $p_5$ and $p_6$ parameters, responsible for the mixture, are however kept fixed 
in the extraction.
The asymmetry between $\bar{u}$ and $\bar{d}$ is parametrized through the ratio:
\begin{equation}
    \frac{\bar{d}}{\bar{u}}(x,Q_0) = p_0x^{p_1}(1-x)^{p_2} + 1 + p_3x(1-x)^{p_4}\,,
\end{equation}
but also all of these parameters are kept fixed.
Finally, we assume that there is no asymmetry between the strange and anti-strange distributions:
\begin{equation}
    (s-\bar{s})(x,Q_0)\equiv0\,.
\end{equation}

The number and momentum sum rules are used to fix the following three normalization parameters: $p^{u_v}_0,p^{d_v}_0$ and $p^{\bar{d}+\bar{u}}_0$. 
Furthermore, instead of fitting the normalization parameters for the gluon and strange distributions we fit the momentum fractions carried by these distributions, referred to as $g_{sum}$ and $s_{sum}$, respectively. The two sets of parameters are directly related. Specifically, for the gluon we have:
\begin{equation}
    p^{g}_0 = \frac{g_{sum}}{\int_0^1\left(\frac{xg}{p^{g}_0}\right)\mathrm{d}x}.
    \label{eq:gluonSumParameter}
\end{equation}
Similarly, for the $s+\bar{s}$ combination we fit its total momenta $s_{sum}$ as a proportionality factor to the remaining momenta after subtracting the momenta of $u_v,d_v$ and $g$ from the total proton momentum,
\begin{equation}
    p^{s+\bar{s}}_0 = \frac{M_{rem}}{3}\frac{s_{sum}}{\int_{0}^1x(s+\bar{s})/p^{s+\bar{s}}_0\mathrm{d}x}\,,
    \label{eq:ssbarSumParameter}
\end{equation}
where $M_{rem} = 1 - M_{u_v} - M_{d_v} - M_{g}$ and $M_i$ is the total momentum fraction carried by flavour $i$.

The heavy-quark PDFs (charm and bottom) are generated perturbatively, through the DGLAP evolution, and do not require parametrizing. 
From the remaining parameters we still fix seven which are not well constrained. This leaves 15 free parameters that are varied during the analysis. The overview of the parameters together with the fixed values is provided in Tab.~\ref{tab:parameterization}.

\begin{table}[t]
    \centering
    \begin{tabular}{c|cccccc}
\textsc{Parameter}	& $u_v$
		& $d_v$
		& $\bar d+\bar u$
		& $\bar d/\bar u$
		& $g$				
            & $s+\bar s$\\ \hline
$p_0$		& --
		& --
		& --
		& 35712
		& \,\,$\texttt{fit}^\dagger$
            & \,\,$\texttt{fit}^\dagger$\\
$p_1$		& \texttt{fit}
		& \texttt{fit}
		& \texttt{fit}
		& 2.15655
		& \texttt{fit}
		& -0.20775\\
$p_2$		& \texttt{fit}
		& \texttt{fit}
		& \texttt{fit}
		& 193.63
		& \texttt{fit}
		& 8.3286\\
$p_3$		& 0.0
		& -3.503
		& 0.0
		& 17.0
		& \texttt{fit}
		& 0.0\\
$p_4$		& \texttt{fit}
		& \texttt{fit}
		& \texttt{fit}
		& 52.2886
		& \texttt{fit}
		& 14.606\\ 
$p_5$       & --
            & 0.0036
            & --
            & --
            & --
            & --\\ 
$p_6$       & --
            & 2.0
            & --
            & --
            & --
            & --\\
\hline
\end{tabular}
    \caption{The fixed and fitted parameters. The dash indicates that the parameter is either fixed through sum rules or not available in the parametrization. ${}^\dagger$Note that we are fixing the total momentum of the gluon and $s+\bar{s}$ instead of their $p_0$ parameters (see \cref{eq:gluonSumParameter,eq:ssbarSumParameter}).}
    \label{tab:parameterization}
\end{table}

\subsection{Markov chain Monte Carlo}
\label{subsec:mcmc}
%
In this section we describe steps involved in our MCMC analysis. This will include the algorithm employed, the processing of the generated chains including autocorrelation estimates, questions about convergence, as well as the setup of the corresponding hyper-parameters.

For the purpose of our analysis we use the adaptive Metropolis-Hastings (aMH) \cite{Haario2001amh} algorithm. This choice is driven by a number of reasons. Firstly, we aim for an algorithm that is straightforward yet versatile and robust. At the same time we need to minimize the number of required $\chi^2$ evaluations which we deem to be the limiting factor in terms of computational cost. The aMH fulfils these requirements and what is more, compared to the standard random-walk Metropolis-Hastings algorithm~\cite{Metropolis:1953,Hastings1970}, it allows to significantly reduce the autocorrelation of the chains which is crucial for overall efficiency.

\subsubsection{Metropolis-Hastings algorithm}

We start by briefly describing the standard Metropolis-Hastings (MH) algorithm to generate Markov chains and later explain the modifications needed in its adaptive incarnation.
A Markov chain is a list of states $\{\x\}$, where the probability of a state $\x_{i+1}$ depends only on the previous state $\x_{i}$. In the MH algorithm a new state, $\x_{i+1}$, is generated through a two step procedure: first, a candidate state $\tilde{\x}_{i+1}$ is proposed through a proposal distribution $q(\tilde{\x}_{i+1},\x_i)$. 
Usually, a MH-like algorithm is named after its proposal distribution,
thus the aMH algorithm defines a specific choice of proposal distribution that is explained below. Second, the candidate state is accepted/rejected according to the acceptance probability
\begin{equation}
a(\x_i,\tilde{\x}_{i+1}) = \min\big(1,r(\x_i,\tilde{\x}_{i+1})\big)
\label{eq:MCMC_MHacceptance}
\end{equation}
where $r(\x_i,\tilde{\x}_{i+1})$ is the Hastings-ratio given by:
\begin{equation}
r(\x_i,\tilde{\x}_{i+1})
= \frac{\pi(\tilde{\x}_{i+1})q(\tilde{\x}_{i+1},\x_{i})}{\pi(\x_i)q(\x_{i},\tilde{\x}_{i+1})}\,.
\label{eq:HastingsRatio}
\end{equation}
The most important ingredient is $\pi(\x)$, the so-called \textit{invariant} distribution. The Hastings-ratio guarantees that the marginal distribution for each state of the Markov chain converges towards $\pi(\x)$ as the number of iterations $i$ increases and the states of the chain can be identified with samples of $\pi(\x)$.
In other words, the distribution of the states follows the invariant distribution $\pi(\x)$ as $i$ goes to infinity and one can extract information about $\pi(\x)$ from the samples, e.g.~by calculating expectation values from the samples. Therefore, for our application we want to set $\pi(\x)$ to the distribution of the fit parameters given the experimental measurements. $\pi(\x)$ can be constructed, using Bayes theorem, as a product of a prior, $p(\{p_k\})$, and the likelihood function, $l(\{p_k\}|D)$, which in our case will be given by \cref{eq:likelihood}:
\begin{equation}
    \pi(\x_i|D)\equiv p(\{p_k\}_i|D) \propto p(\{p_k\}_i)\; l(\{p_k\}_i|D).
    \label{eq:Lprior}
\end{equation}
Note that we have made the identification $\x_i\equiv\{p_k\}_i$, \textit{i.e.} at each step $i$ in the Markov chain $\x_i$ is a 15-dimensional vector of our fit parameters $\{p_k\}$.
Also, since $\pi$ will always enter through the Hastings-ratio, \cref{eq:HastingsRatio}, its normalization will not be important.

It should be highlighted that the candidate state will be accepted as the next state in the chain ($\x_{i+1}=\tilde{\x}_{i+1}$) only if it survives the acceptance probability. If the state is rejected the algorithm stays at the same point ($\x_{i+1}=\x_{i}$). This means that if the algorithm is in state $\x_{i}$ and new states are being rejected multiple times the state $\x_{i}$ is repeated the same number of times. This is quite different from standard Monte Carlo rejection algorithms, used in particle physics, where rejected points are discarded without changing the multiplicity of the current state.

In the particular case considered here, we will use a proposal function given by a multivariate Gaussian distribution:
\begin{equation} 
q(\tilde{\x}_{i+1},\x_i) = \mathcal{N}(\x_i,C_0)
\end{equation}
with a covariance $C_0$.
This choice is symmetric, $q(\tilde{\x}_{i+1},\x_i)=q(|\tilde{\x}_{i+1}-\x_i|)$, which ensures that any region of the parameter space can be reached regardless of the starting point. And since the proposed state is given by a random increment around the current state, this choice belongs to the class of random-walk Metropolis-Hastings (RWMH) algorithms. However, the efficiency (in terms of how many steps in the chain one has to generate in order to estimate a quantity to a given precision) of this proposal algorithm strongly depends on the user-set covariance $C_0$. This is a very challenging problem, especially for a large number of dimensions (fit parameters).

\subsubsection{Proposal algorithm: Adaptive Metropolis-Hastings}
\label{subsubsec:algorithm}

Having introduced the standard MH algorithm we now describe the adaptive Metropolis-Hastings algorithm. It consists of three steps:
\begin{enumerate}
    \item Use the RWMH algorithm until $N_0$ samples have been obtained. This includes using the multivariate Gaussian proposal distribution, $\mathcal{N}(\x_i,C_0)$, with the fixed covariance matrix, $C_0$.
\end{enumerate}

\noindent At this stage of the algorithm the sampling of $\pi$ will generally be very inefficient since setting the individual values of the covariance matrix by hand is very tedious. And, even if a well-working $C_0$ is found it might become inefficient in other regions of the available parameter space. Instead the goal at this stage is to accumulate enough samples such that the estimation in the next step become reasonably accurate.

\begin{enumerate}
    \item[2.] Switch to a self learning proposal distribution:
    \begin{equation}
    \begin{split}        
        q(\tilde{\x}_{i+1},\x_i) &= (1 - \beta)\; \mathcal{N}(\x_i,s_d\cdot\overline{C}_{i})
        \\& + \beta\;\mathcal{N}(\MCMCparam_i,C_0),
    \end{split}
    \end{equation}
    where $\overline{C}_{i}$ is a self learned covariance matrix computed from the so far collected samples. The parameter $0<\beta\leq1$ controls the impact of the {\em learned} proposal, and $s_d$ is a dimension-dependent scale parameter introduced to optimise the proposals, usually $s_d= (2.4)^2/d$ with $d$ being the dimension~\cite{Roberts:2001}.
\end{enumerate}

\begin{figure*}[hbt]
    \centering
    \includegraphics[width=0.99\textwidth]{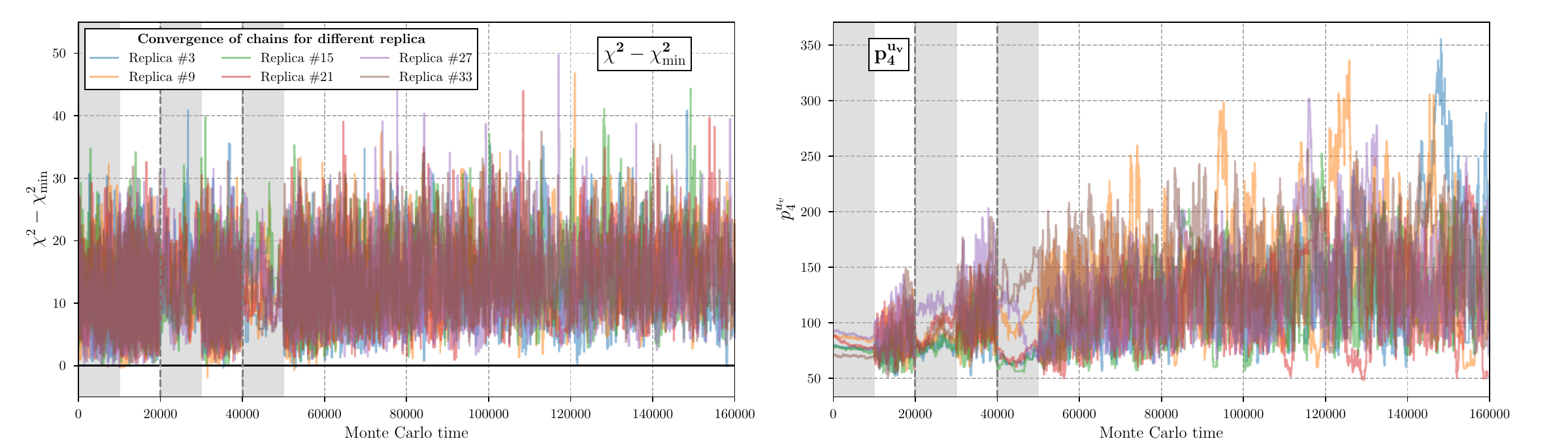}
    \caption{The convergence of six of the 36 generated chains towards the same minimum. The left plot shows the $(\chi^2-\chi^2_{min})$-value for each sample and the right plot the parameter $p_4^{u_v}$ versus the Monte Carlo time. The dashed lines indicate the times when a reset was performed and the grey regions the steps when the static/fixed proposal distribution was used. Note that only the first 160\,000 points are shown (for the remaining 319\,000 points the behaviour is similar to the last shown part).}
    \label{fig:convergence_of_replica}
\end{figure*}

\noindent The self-learned covariance matrix will be updated (or adapted) at each step with the newest sample. This operation can be performed in constant time, since averages and covariance matrices obey a recursion relation.%
\footnote{At each step, $i$, the self-learned covariance is updated through:
    \[ 
    \overline{C}_{i+1} = \frac{1}{i}\left( S_{i+1}-(i+1)\mub_{i+1} \mub_{i+1}^T \right),
    \]
    with $S_{i+1}=\sum_{k=0}^{i}\x_k\x_k^T = S_{i}+\x_{i+1}\x_{i+1}^T$. The mean of samples, $\mub_{i}$, can also be written recursively as $\mub_{i+1}=\mub_{i}+\frac{1}{i+1}(\x_{i+1}-\mub_{i})$.}
It is to be noted that this algorithm technically breaks the Markovian aspect of the generated chain: since the proposal algorithm uses all previous samples, the proposed sample does not depend only on the previous sample any more. It can however be shown that the aMH retains the convergence properties of the MH algorithm under mild conditions~\cite{Roberts:2007}. One of those is to keep $\beta>0$. \cref{fig:convergence_of_replica} shows a trace plot of the final chain, where the gray regions indicate proposals from a fixed covariance matrix from step 1. One can clearly see that switching to the self learned $\overline{C}_{i}$ improves the mixing (the variance between two consecutive states) considerably.

\begin{enumerate}
    \item[3.] Reset the self learned proposal distribution at early stages of the chain.
\end{enumerate}

\noindent  In our trial runs it has proven beneficial to reset the correlation matrix at several points to reduce the impact of the starting point of the chain. This can be easily understood since the correlations among the parameters are generally local in parameter space, such that resetting the estimates when the chain still drifts towards the region of maximum probability removes unwanted information in the estimation of $\overline{C}_{i}$.

\subsubsection{Prior}
\label{subsubsec:prior}

Within the Bayesian approach the construction of an invariant distribution naturally involves the use of prior distributions, see Eq.~\eqref{eq:Lprior}. The prior distribution encodes our initial knowledge about the parameters which can either originate from earlier measurements or from theoretical constraints. 
In our analysis we prefer not to introduce such prior knowledge on the free PDF parameters such that we only use the information coming from the experimental measurements introduced via the likelihood function.%
\footnote{Implicitly we make use of priors for the parameters that we do not fit e.g.~by enforcing the momentum and number sum rules on the PDFs or fixing them to specific values.}

There will be however one exception to this, the ${p_4^{d_v}}$ parameter. In case of this parameter the relatively weak constraints combined with certain shortcomings of our PDF parametrization result in effective independence of $d_v(x,Q_0)$ on ${p_4^{d_v}}$. Specifically, when ${p_4^{d_v}}$ becomes very large it can be absorbed into the ${p_0^{d_v}}$ when using the number sum rule to normalize this distribution,\footnote{In principle this shortcoming is present for all $p_3$ and $p_4$ parameters. However, absorbing the parameter into the normalization comes with a change of shape and reduces flexibility. Thus if the distribution is sufficiently well constrained this shortcoming is irrelevant.} see Eq.~\eqref{eq:parmdv}.

This poses a serious problem: one of the key conditions for a MCMC algorithm to converge is that the target distribution is a proper probability distribution. If $\pi$ becomes independent of ${p_4^{d_v}}$ in the limit of the parameter going to infinity, $\pi$ is non-normalizable and thus not a proper probability distribution~\cite{Jones:2022}.

One solution, commonly used for weakly constrained parameters, is to simply keep the parameter fixed. This has the additional, often desired, effect of removing the non-Gaussian correlations usually introduced through the loosely constrained parameter, which would make the uncertainty estimation very challenging with traditional methods. But, fixing the parameter typically decreases the available space for the other parameters. A MCMC setup however, is well equipped to handle such non-Gaussian correlations and we can use $p_4^{d_v}$ to confirm this capability of the method.

For this reason we opt to modify our target distribution $\pi$, and introduce a prior for this parameter instead of fixing it to a specific value. This should be done carefully in order to not distort the sampling of the $\chi^2$ function, since we want to keep the correspondence between $\chi^2$-value and goodness-of-fit intact. We use an uniform (or flat) prior $U(p_4^{min},p_4^{max})$, which is constant inside the interval $[p_4^{min},p_4^{max}]$ and zero outside. A uniform prior does not move the mode of the likelihood function and therefore the results stay unbiased as long as we restrict $p_4$ to the correct region; but still sets bounds on the parameter. 

\begin{figure}[hbt]
    \centering
    \includegraphics[width=0.49\textwidth]{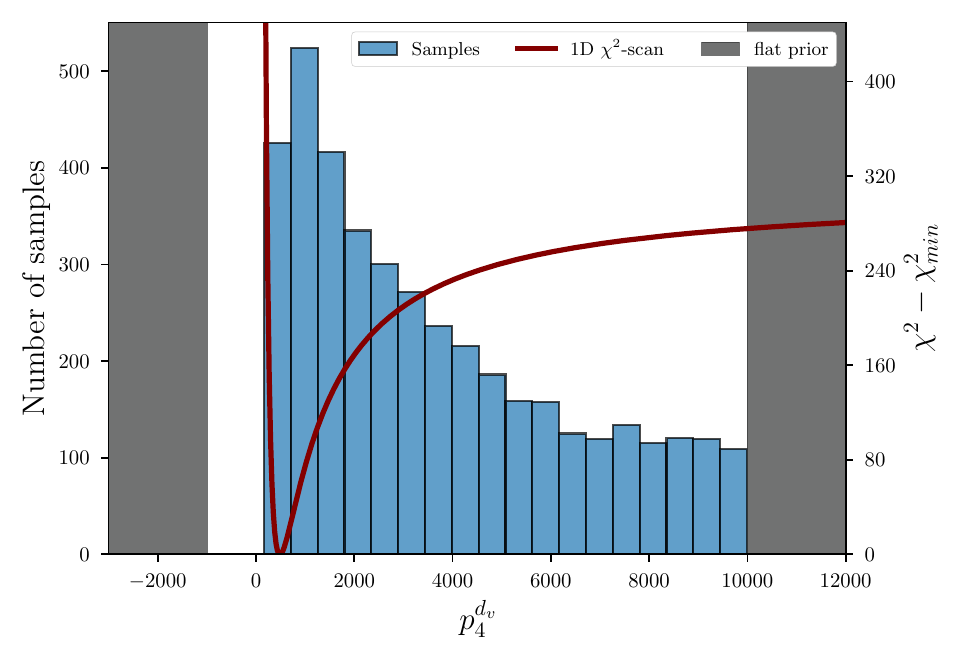}
    \caption{The samples/marginal distribution for the parameter $p_4^{d_v}$ (left axis) along with a one-dimensional $\chi^2$-scan (right axis). The gray-shaded areas indicate the regions excluded by the flat prior.}
    \label{fig:dv_p4_vs_chi2}
\end{figure}

Having analysed samples of the $p_4^{d_v}$ parameter and the behaviour of the one dimensional scans of the $\chi^2$ function (with other parameters fixed to values close to the minimum) we fix the bounds of the prior to $p_4^{min} = -10^3$ and $p_4^{max} = 10^4$, as displayed in Fig.~\ref{fig:dv_p4_vs_chi2}. Note that the 1D $\chi^2$-scan shows an increase of $\sim300$ points by the time $p_4^{d_v}$ reaches the upper limit of the prior. Naively one would assume that these large parameter values are safely excluded from the experimental data. However, this $\chi^2$-difference can easily be compensated for by the other 14 fit parameters.

\subsubsection{Starting points}
\label{subsec:startingpts}
Formally the final Markov chain is independent of the starting point. In practice the starting point significantly impacts the efficiency of the algorithm, especially for the proposal algorithm we opted for. For this reason we follow the recommendation of Ref.\,\cite{Jones:2022} and select the initial points to be close to the global minimum found by a traditional gradient decent minimization algorithm. To be more specific we performed three minimizations using the BFGS-algorithm~\cite{Broyden1970tco,Fletcher1970ana,Goldfarb1970AFO,Shanno1970COQ} starting from different initial points and selected the minimum with the lowest $\chi^2$ ($\chi^2_{\mathrm{min}}/\rm{(data\,points)}=2366.03/1984$). Then we performed one-dimensional $\chi^2$ scans to estimate the variance for each fit parameter, and when choosing the starting point for each individual chain all the parameters were displaced randomly according to these variances. 
Altogether 36 independent Markov chains were generated and the variation of the $\chi^2$ of the starting points from the ``BFGS minimum'' were: $\chi^2_{\mathrm{start}}-\chi^2_{\mathrm{min}}\in [16,1296]$ points, which corresponds to  0.7\% to 55\% of $\chi^2_{\mathrm{min}}$. 
Finally, for each of the 36 generated chains a random number generator~\cite{sanderson2016armadillo,sanderson2019practical} was initialized with different seeds.

Note that this choice of starting points led to all chains converging towards the same minimum as the one found by the BFGS-algorithm. In trial runs, where the chains started from random points in the parameter space, the chain often got ``stuck'' in local minima with much larger $\chi^2$-values.

\subsubsection{Hypersetup}

As already mentioned in the previous section we produced 36 independent Markov Chains which we later process and combine into a single set of samples. The main motivation for this is the trivial parallelization by running multiple chains at the same time, yielding independent samples with negligible computational overhead. Of course, each chain needs to be independent and sufficiently long to thermalize and collect enough samples. Additionally, from a general point of view, having multiple chains starting from different initial points potentially increases the chance of identifying better minima, if they exist. This enhances our confidence that our minimum is representative and not an artifact of limited sampling.

The parameters of the adaptive Metropolis Hastings algorithm are chosen as follows:
\begin{itemize}
\item $C_0$ is diagonal with values given by $0.005\sigma_{p_i}$, where $\sigma_{p_i}$ were estimated from one-dimensional $\chi^2$-scans along $p_i$ parameters at a minimum estimated with a traditional minimization algorithm (as described above). They correspond roughly to a distance needed for a $\chi^2$-increase of 100 points along their respective scans.
\item The number of points in the initial phase is set to $N_0=10\,000$. This choice reflects the need for sufficient statistics to reliably initialize the self learning phase.
\item The parameter responsible for mixing of the fixed and self learned covariance matrices in the proposal function is set to $\beta=0.1$. This value was chosen following the recommendation of~\cite{Haario2001amh}.
\item After 20\,000 and 40\,000 points the self learned covariance is reset. This means that the true sampling with the aMH algorithm begins after 50\,000 points. For each reset we modify the fixed covariance matrix $C_0$ to 5\% of the diagonal values of the last estimate of $\bar{C}_i$.  
\end{itemize}

Each of the 36 chains has a length of 479\,000 samples giving in total $N_{\text{tot}}=17\,244\,000$ samples generated in 14 days on 36 cores\footnote{Further speed up may not be possible as each chain must have a certain minimum length, e.g.~to ensure thermalization (see below) and collect enough samples to provide sufficient statistics.} on the Palma II cluster. Having such long chains ensures that they are very well converged. This allows us to remove autocorrelations, whilst still having enough samples to estimate derived quantities confidently.

\begin{figure}[hbt]
    \centering
    \includegraphics[width=1\linewidth]{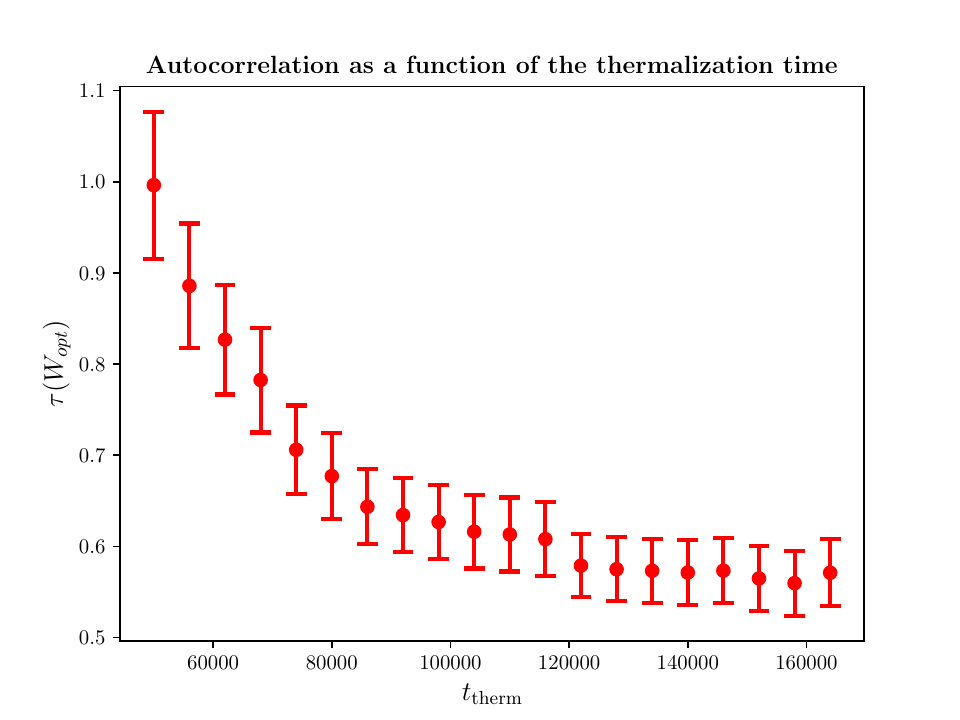}
    \caption{The integrated autocorrelation time $\tau_{\rm int}$ as a function of the thermalization time $t_{\rm therm}$ (after thinning with $\eta=3\,000$, motivated in the next section). At $t_{\rm therm}>120\,000$ the autocorrelation time exhibits a plateau. Repeating this exercise for each parameter individually, we find the same plateau starting in the range of $t_{\rm therm}\in[50\,000,130\,000]$.}
    \label{fig:tau_as_function_of_t_therm}
\end{figure}

We are very conservative in estimating the so-called thermalization time $t_{\text{therm}}$ (the number of samples removed from the start of the chain). In our analysis we estimate the chains to have thermalized after 140\,000 steps, and we remove these samples from the chains before further analysis. This choice can be best understood by considering the integrated autocorrelation time $\tau_{\text{int}}$ of the entire chain as a function of the thermalization time, see \cref{fig:tau_as_function_of_t_therm}. The autocorrelation time is a measure of how dependent each step is on its predecessors. $2\tau_{\text{int}}$ can be interpreted as the average number of steps one has to take to arrive at the next uncorrelated sample, thus its optimal value is $\tau_{\text{int}} = 1/2$. The estimation of $\tau_{\text{int}}$ is based on the $\Gamma$-method~\cite{Wolff:2003sm}, summarized in \cref{app:GammaMethod} (see \cref{eq:tau_int}). This is the primary use-case for $\tau_{\text{int}}$ which we will utilize in the following section. A secondary use-case is its sensitivity to when the chain drifts in parameter space, i.e.~$\tau_{\text{int}}$ rises if the chain drifts. Bias originating from the dependence on the starting point presents itself by a drift towards a region of better likelihood. 
After removing $\sim 120\,000$ steps almost no remaining dependence on the starting point is found as indicated by the plateau in the figure. Similar estimates for each parameter individually give the same plateau for $t_{\text{therm}}\in[50\,000,130\,000]$. 
Additionally, looking directly at the time series displayed in \cref{fig:convergence_of_replica} (where we show only the first 160\,000 points) provides a common-sense way to make sure that the estimates based on the autocorrelation time are not unreasonable and that there is no obvious dependence on the initial point. Of course, such a ``by-eye'' strategy can not tell us precisely where to set the burn-in time but can verify if there is no obvious fault in the value we chose (e.g.~by spotting regions of huge correlations seen around time $\sim40\,000-50\,000$).

Having removed samples before the thermalization time 12\,204\,000 samples remain.

\subsubsection{Thinning}

By nature of the MCMC algorithm it yields autocorrelations in the generated chains. In principle, one can use such autocorrelated samples directly, however, from a practical perspective it is easier to remove the autocorrelation from the individual chains (which we call replicas\footnote{This is not to be confused with the term replica in the MC replica method of NNPDF~\cite{Forte:2002us,DelDebbio:2007ee}. In the terminology used in MCMC inference (and this work) replicas refer to individual chains that were generated in a statistically independent manner.} in the following). There are a number of advantages in doing so. First of all, having autocorrelation-free (and independent) replicas allows combining them into a single, final set of samples in a straightforward way. Further, the lack of autocorrelation allows us to use simple estimators, e.g.~when computing uncertainties based on the samples making the result much more user-friendly. Finally, as we will see below this considerably reduces the number of samples, in turn lowering the computational cost of performing any operations on the chain. 

In order to remove correlations we will carry out thinning~\cite{Jones:2022} on each of the replicas (chains) by a factor $\eta$. Specifically, for each replica, $r$:
\begin{equation}
    \{\x_1,\x_2,\x_3\dots,\x_{N_r}\}^r
\end{equation}
we will subsample the chain keeping only every-$\eta$ point.%
\footnote{At this stage we use chains for which thermalization phase has been already removed.}
Additionally, we introduce an offset $\delta<\eta$ which ensures we keep the points with the lowest values of $\chi^2$ in the thinned chain, resulting in:
\begin{equation}
\{\x_{\delta},\x_{\eta+\delta},\x_{2\eta+\delta},\dots,\x{_{(N_r/\eta-1)\eta+\delta}}\}^r \;.
\end{equation}
We choose here to use a common value of the thinning rate, $\eta$, and offset, $\delta$, for all considered chains but this can in principle be adjusted for individual replicas.
After thinning the final number of independent samples is $\sum_r N_{r}/\eta = N/\eta$. 

Now the question is what the right value of $\eta$, which will ensure that the chains are indeed free of autocorrelations, is. At the same time $\eta$ needs to be small enough so that the final sample size is large enough for the central limit theorem to apply.
In order to estimate the value of the thinning rate we will use the integrated autocorrelation time $\tau_{\text{int}}$ from the previous section (computed by the $\Gamma$-method~\cite{Wolff:2003sm}). As already explained, $\tau_{\text{int}}$ indicates the average time to find the next sample that is uncorrelated from the current sample, giving us a proxy for the thinning rate ($\eta\sim2\tau_{\text{int}}$). If the sample is free from autocorrelations we should find $\tau_{\text{int}}\approx \frac{1}{2}$. In practice, we will thin a chain with $\eta>2\tau_{\text{int}}$ and recompute $\tau_{\text{int}}$ for the thinned chain. If the newly computed $\tau_{\text{int}}\approx \frac{1}{2}$ we stop, if $\tau_{\text{int}} > \frac{1}{2}$ we will increase $\eta$ and continue with this procedure until we obtain $\tau_{\text{int}}\simeq \frac{1}{2}$.%
\footnote{If we use $\eta\gg2\tau_{\text{int}}$ we should remove all autocorrelations and the thinned chain should have $\tau_{\text{int}}\approx \frac{1}{2}$. In practice, the $\Gamma$-method assumes the autocorrelation to be purely exponential, however, there are other contributions to the autocorrelation which scale as polynomials (i.e.~slower decay) and are poorly estimated.}

\begin{figure*}[htb]
    \centering
    \includegraphics[width=0.33\textwidth]{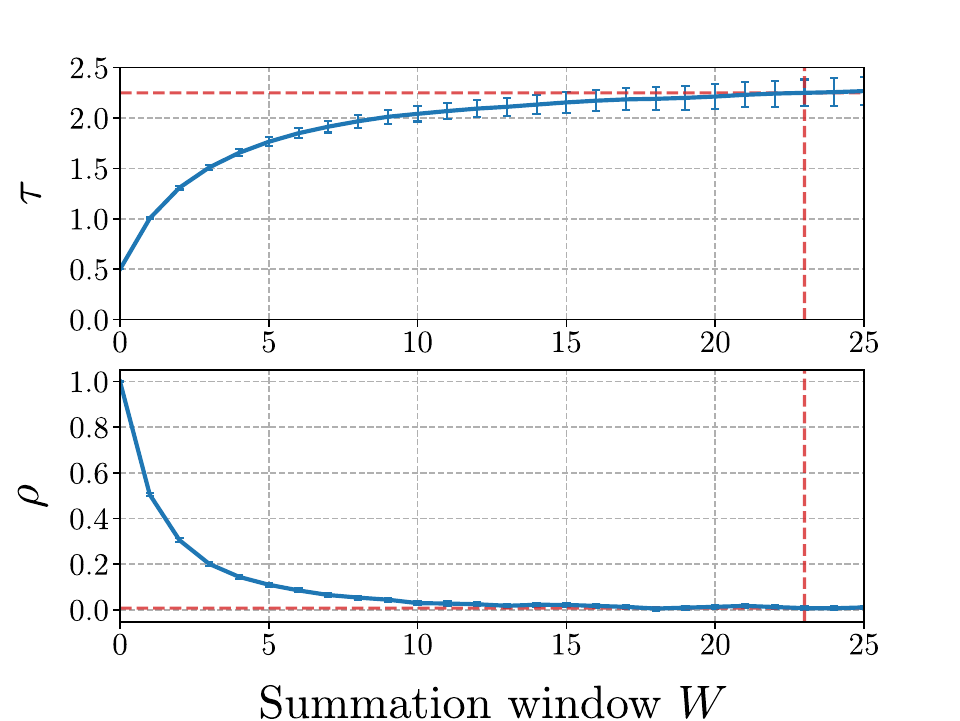}%
    \includegraphics[width=0.33\textwidth]{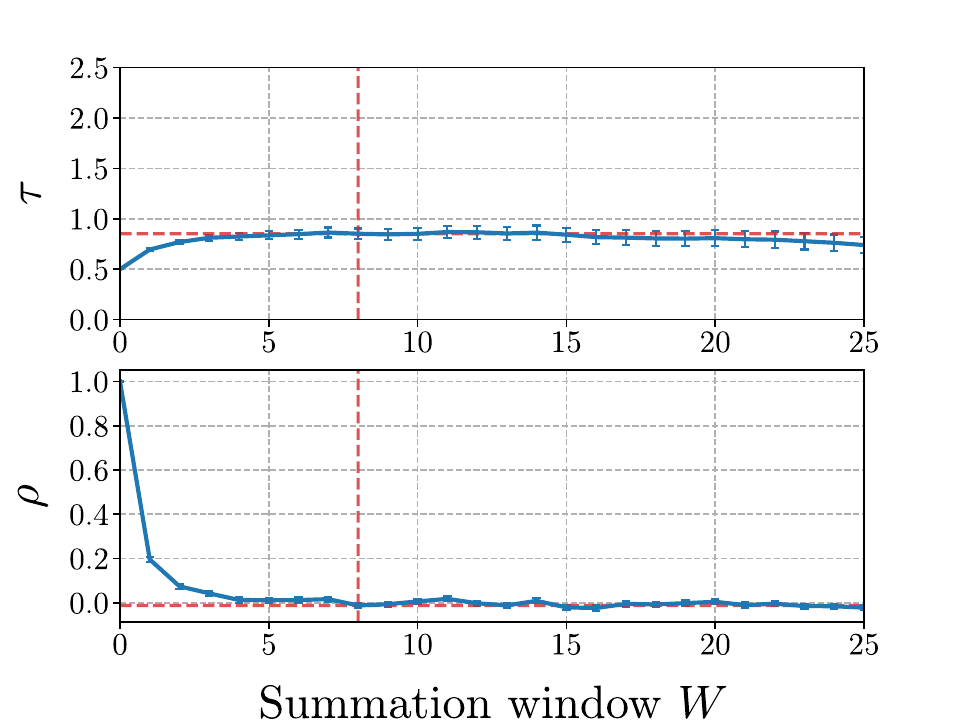}%
    \includegraphics[width=0.33\textwidth]{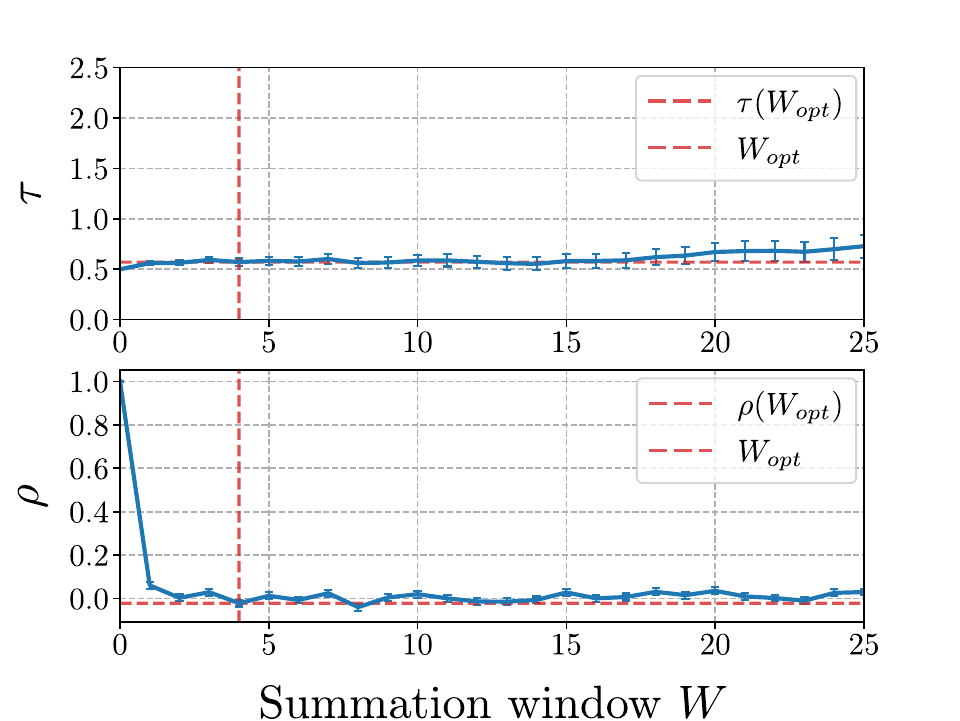}
    \caption{The integrated autocorrelation time $\tau_{\rm int}(W)$ (upper plot, see \cref{eq:tau_int}) and the autocorrelation function $\rho(W)$ (lower plot, see \cref{eq:rho_W}) as a function of the summation window $W$ in units of Monte Carlo time. In principle $\tau_{\rm int}$ is defined by summing over $\rho(W)$ until infinity. Since this is impossible we resort to the $\Gamma$-method that defines an algorithm to find the optimal value of $W$ to truncate the summation. This is referred to as the optimal summation window, $W_{\rm opt}$, indicated by the vertical red dashed lines. The horizontal red dashed lines show the estimated $\rho(W_{\rm opt})$ and $\tau_{\rm int}(W_{\rm opt})$. Going from left to right we have thinned the original chain with factors $\eta=\{500,1500,3000\}$ and thus reducing $\tau_{\rm int}$.}
    \label{fig:autocorrelation_vs_thinning}
\end{figure*}

We moved the description of the $\Gamma$-method to \cref{app:GammaMethod}. Here we present how the method works in practice when thinning chains with different rates and computing the correlation time. In Fig.~\ref{fig:autocorrelation_vs_thinning} we show the impact of thinning with different rates on one of the generated chains (similar results are obtained for all involved chains). We display results for three thinning rates $\eta=\{500,1500,3000\}$, one can see that this indeed leads to a systematic reduction in autocorrelation time and for $\eta=3000$ we obtain $\tau_{\text{int}}=0.57 \pm 0.035$. Further thinning does not reduce $\tau_{\text{int}}$. We do not show it here but the initial chain has $\tau_{\text{int}}\approx 500$.  

Finally, we set $\delta=671$ by hand. This choice is entirely motivated by setting the offset to include the sample with the best $\chi^2$-value in the final set of samples.
With the final thinning rate of $\eta=3000$ (and removing the thermalization phase of 140\,000 points) we end up with $N_{r,\text{thin}}=113$ uncorrelated points per replica. This way the replicas still contain enough samples so that we do not need to worry about introducing bias due to large statistical errors and we can safely combine the individual 36 chains to form the final combined sample of $N=4068$ uncorrelated points. This sample will be used to perform the final analysis and extract PDFs. From this point, we assume the chain to be free from autocorrelation.

We should also note that it can be shown that the variance of a thinned Monte Carlo estimator is always larger than the variance of the original Monte Carlo estimator~\cite{Jones:2022}. In other words we are loosing statistics by thinning and are better off using the full chain instead. However, from a practical point of view our sample size is large enough even after thinning.
Furthermore, using the fully correlated samples to compute uncertainties of PDF-dependent observables is prohibitive simply due to the required computational effort.
Thinning is actually common practise for large autocorrelation times, e.g., in lattice QCD. And in fact having $N=4068$ different PDF sets for the uncertainty estimation is still by far the largest set for uncertainty estimation available in the literature.

\begin{figure*}[t]
    \centering
    \includegraphics[width=\textwidth]{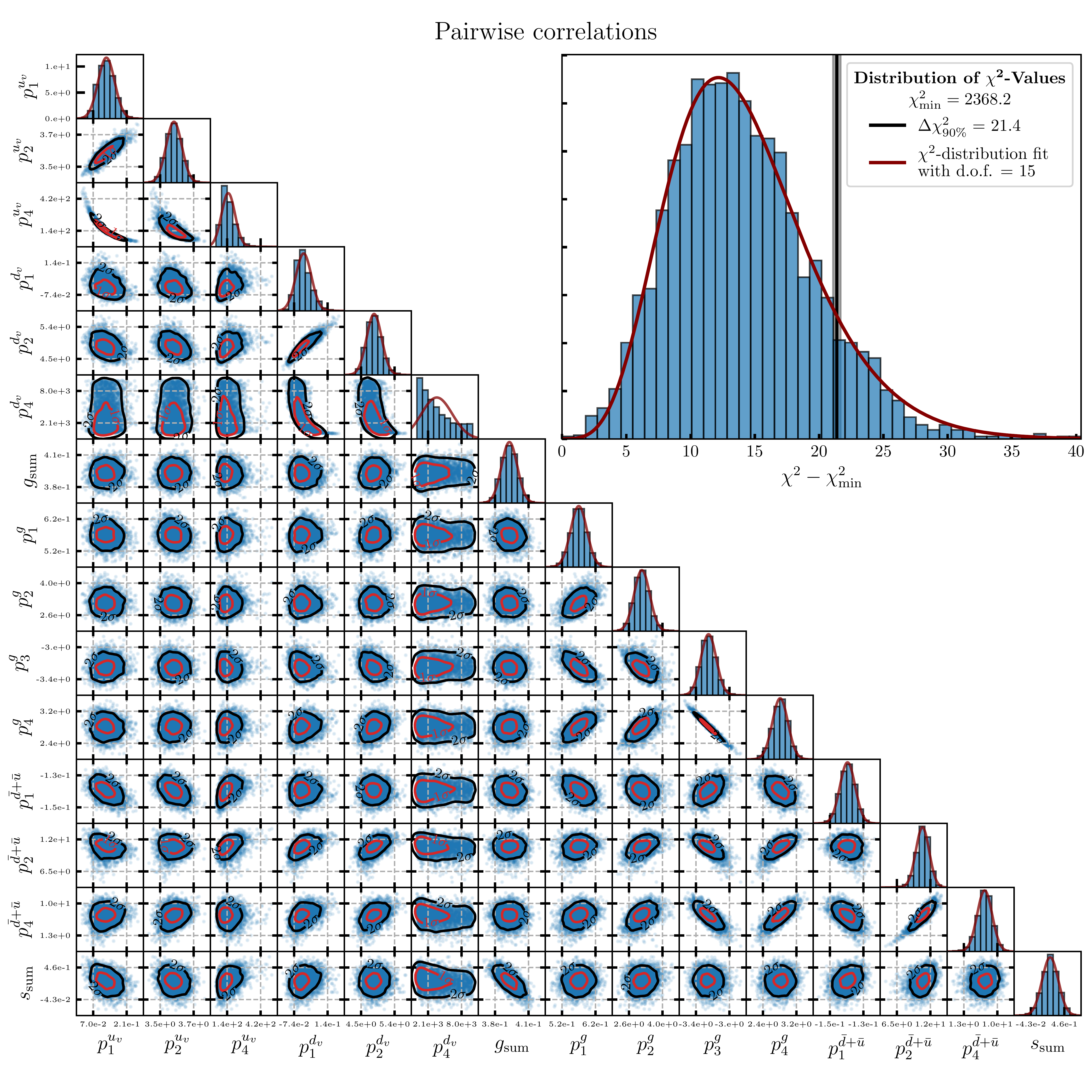}
    \caption{The pairwise distribution of parameters. The top right figure shows a histogram representation of the $\chi^2$ values (blue bars), a $\chi^2$-distribution fit to the histogram with fixed $\rm{d.o.f.}=15$ (red curve) and a 90\%-quantile estimation (black horizontal line). The smaller plots on the lower left show various marginal distributions. On the diagonal we show the 1D distributions for each parameter as a histogram (blue bars) with a fitted normal distribution (red lines). The scatter-plots show the 2D distributions for each parameter pair (blue points). The red and black curves indicate the regions for $1\sigma$ and $2\sigma$ of the number of samples based on kernel density estimations.}
    \label{fig:pairwise}
\end{figure*}

To visualize the final sample, in \cref{fig:pairwise}, we show pairwise distributions for all combinations of free PDF parameters together with the distribution of the $\chi^2$ values. This will be further useful when analysing the final sample in Sec.~\ref{sec:analysis}.

\section{Analysis}
\label{sec:analysis}

In the previous section we obtained an uncorrelated set of samples in PDF parameter space each corresponding to an independent PDF set which can now be used to calculate central predictions and uncertainties of PDFs and PDF-dependent quantities. There are different ways of how uncertainties can be constructed from the samples. 

In general the best practice is to evaluate the desired observable $\Ob$ for each sample and plot the distribution of results, e.g.~as a histogram for 1D observables or as corner plots for multidimensional observables. This distribution corresponds to the projection of the experimental measurements and uncertainties to the observable of interest, since we have
\begin{align}
    \{p_k\}_i \sim p(\{p_k\}|D) \quad\Rightarrow\quad \Ob(\{p_k\}_i) \sim p(\Ob|D) .
\end{align}
This is the most important feature of the MCMC-method; the samples can be directly used to infer the posterior distribution of any observable. 

As a showcase on how powerful these figures can be, we return to the marginal distribution of the $\chi^2$-values, the marginal distributions of all parameters and their pairwise correlations in a corner plot in \cref{fig:pairwise}. The large plot in the top right corner is a histogram representation of the $\chi^2$-values for all samples in blue. The black vertical line at $\chi^2-\chi^2_{\rm min}=21.4$ is a 90\% quantile estimation~\cite{Doss2014} (with a respective 95\% uncertainty band given in gray) on the distribution signifying that 90\% of all $\chi^2$ values have a $\chi^2$-increase less than 21.4 points. The red curve is a fit to a $\chi^2$-distribution with fixed $\rm{d.o.f.}=15$. 
The fact that the curve follows the histogram closely indicates that the posterior distribution of our parameters close to the minimum indeed behaves like a $\chi^2$-distribution with the same number of degrees of freedom, ultimately showing that our assumptions about the theory and experimental data are justified. However, it is worth noting that an ideal $\chi^2$-distribution with $N_{\rm{d.o.f.}}=15$ would have its 90\% quantile at an increase of 22.3~\cite{Wilks:1938dza}. The diagonal plots are the marginal distributions of the parameters (as a histogram in blue), where the red curve is a fit to a 1D Gaussian to easily guide the eye to which parameters are behaving non-Gaussian. Finally, the scatter-plots show the marginalized 2D distributions for each parameter pair. The red and black lines indicate the regions of higher/lower density based on kernel density estimations~\cite{Scott:1992}. From this set of plots numerous non-trivial results about the parameters can be inferred, including that the parameter that shows the most significant deviation from Gaussianity is the $p_4$ parameter, and that interestingly the full set of gluon parameters ($\{g_{\rm sum},\dots,p^g_4\}$) is very close to a Gaussian including their correlations. No new calculations had to be performed to obtain this plot, since all of the information is already contained in the chain.

Nevertheless, in usual applications the direct use of samples to compute distributions for derived quantities is not that common, and for practical reasons it is useful to also be able to reduce the obtained information to the commonly used central value and an upper and lower bounds.

\subsection{MCMC uncertainty estimates}
\label{subsec:errors}

Here we present three different options to define the central value with its bounds and compare among them as well as to confront them with the uncertainties obtained via the traditional Hessian approach of \cref{subsec:Hess}.\footnote{Note that a comparison with the Monte Carlo replica method (cf. \cref{subsec:MCrep}) would be interesting as well, especially with respect to the apparent deformations in the likelihood. However this comparison is beyond the scope of this paper.}
For each approach we define 90\% confidence intervals $[\Ob_-,\Ob_+]$ for any PDF-dependent observable, $\Ob$, alongside with a ``central'' value $\Ob_*$. For the following it is convenient to define the shorthand notation $\Ob_i=\Ob(\x_i)$. Further, we define the sample with the lowest $\chi^2$-value as $\x_{\text{min}}$.

These options are: 
(i) \asym, 
(ii) \aasym,
(iii) \cumchi.

\subsubsection{$\alpha$\%-symmetric}
\label{subsubsec:alphaSym}
This method assumes that the distribution of $\Ob$ is Gaussian and simply computes the central value as mean and standard deviation leading to a symmetric confidence interval:
\begin{subequations}
\begin{align}
	\Ob_* &= \langle \Ob \rangle\\
	\Ob_- &= \Ob_* - 1.645\,\sigma_{\Ob}\\
	\Ob_+ &= \Ob_* + 1.645\,\sigma_{\Ob} \,,
\end{align}
\end{subequations}
with
\begin{subequations}
\begin{align}
	\langle \Ob \rangle &= \frac{1}{N}\sum_i^N\Ob_i,\\
	\sigma_{\Ob} &= \sqrt{\frac{1}{N-1}\sum_i^N\left(\Ob_i-\langle \Ob \rangle\right)^2},
\end{align}
\end{subequations}
and $N$ being the number of PDFs in the sample, and the factor 1.645 accounts for the fact that we use 90\% confidence instead of 68\%.

\subsubsection{$\alpha$\%-asymmetric}
\label{subsubsec:alphaAsym}
In this method we do not assume a Gaussian distribution of the samples~\cite{Gbedo:2017eyp}. We define the central value as the set of parameters that has the lowest $\chi^2$-value, $\Ob_*=\Ob(\x_{\text{min}})$. The potentially asymmetric confidence interval is given by 90\%-quantile estimations on the upper and lower bound around the central value. To compute the quantiles we order the values of the observable by value and divide the samples into two sets: $\Ob_{\text{lower}}=\{\Ob_i|\Ob<\Ob_*\}$ and $\Ob_{\text{upper}}=\{\Ob_i|\Ob>\Ob_*\}$. The limits on the (asymmetric) confidence interval are defined as the 10\%-quantiles from $\Ob_{\text{lower}}$ for the lower bound and as the 90\%-quantile from $\Ob_{\text{upper}}$ for the upper bound~\cite{Gbedo:2017eyp,Jones:2022}.

\subsubsection{Cumulative $\chi^2$}
\label{subsubsec:cumchi}
In this method the confidence intervals are also defined as 90\% quantiles but now they are based on the values of the $\chi^2$ function for each sample~\cite{Putze:2008ps}. We perform a 90\%-quantile estimation on the distribution of $\chi^2$-values and define the bound as the minimum/maximum reach of the observable found within this 90\%-quantile. 

The bounds, this method places, can be intuitively understood as the maximum reach an observable can have whilst still being within a 90\%-level agreement with the experimental data sets used for the PDF extraction.

\subsubsection{Comparison}
\label{subsubsec:compUncer}
\begin{figure*}[htb!]
	\centering
    \includegraphics[width=0.33\textwidth]{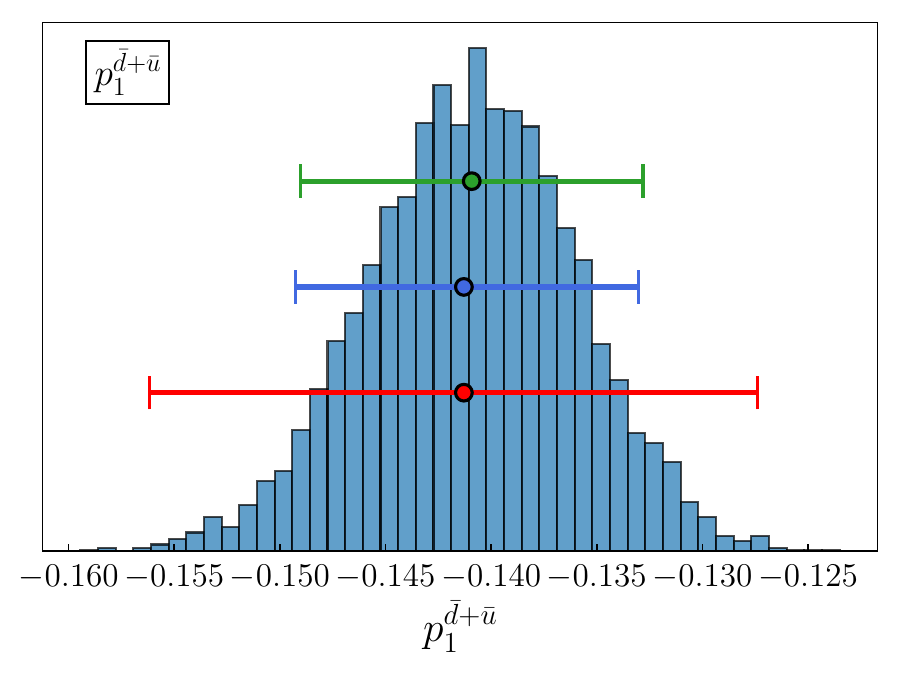}%
    \includegraphics[width=0.33\textwidth]{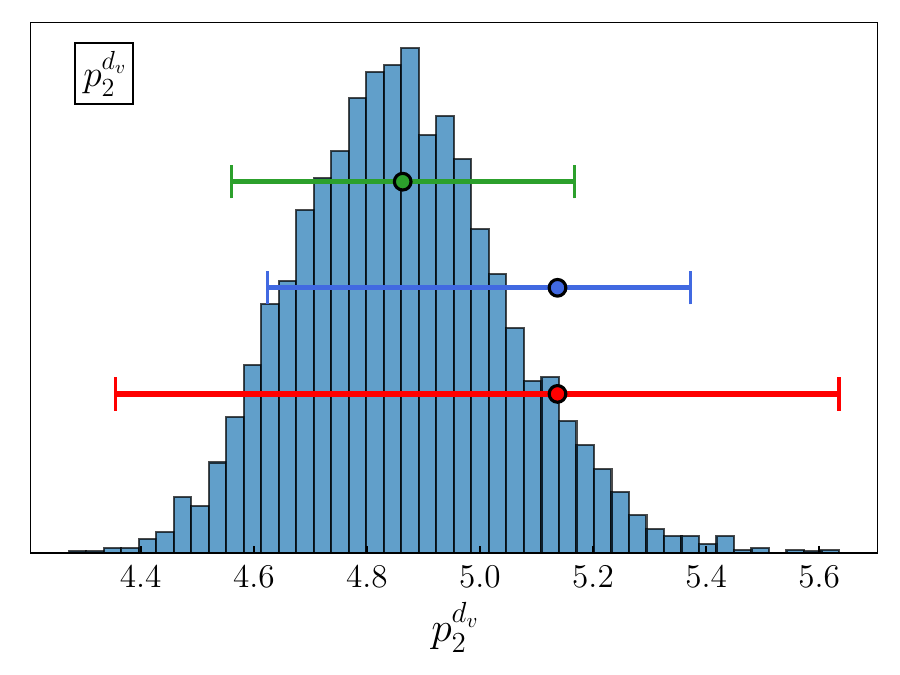}%
    \includegraphics[width=0.33\textwidth]{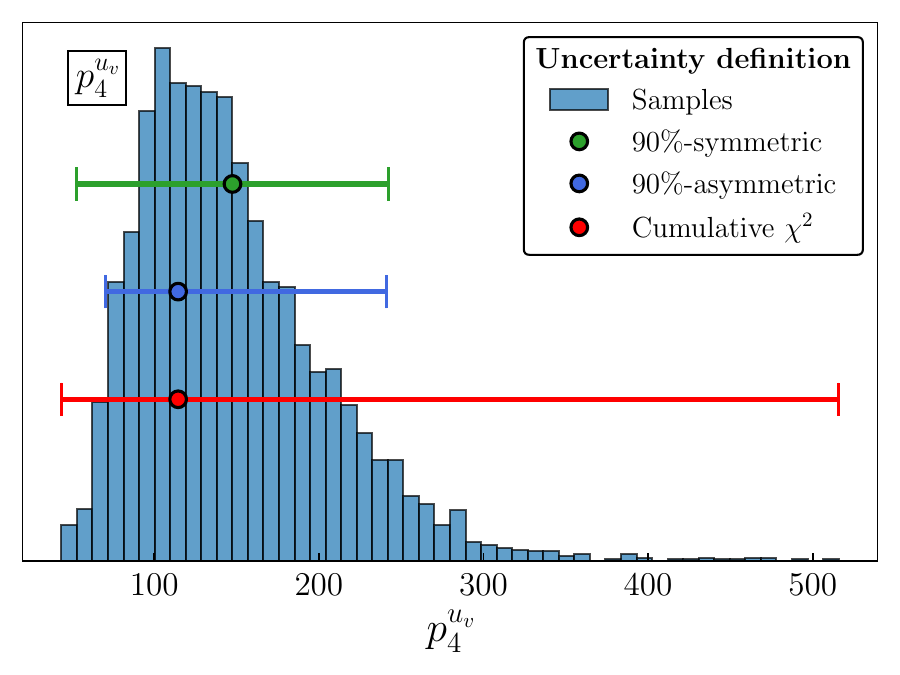}%
	\caption{Marginal parameter distributions for the parameters $p^{\bar{d}+\bar{u}}_1$ (left), $p^{d_v}_2$ (center) and $p^{\bar{u}}_4$ (right). For each figure the three uncertainty estimations are shown: 90\%-symmetric (green), 90\%-asymmetric (blue) and cumulative $\chi^2$ (red).}
	\label{fig:marginal_param}
\end{figure*}

In order to compare the above three methods for propagating uncertainties we first look at individual parameters defining PDFs and then at the PDFs themselves. To look at the individual parameters we compute the marginal distribution for each of the varied parameters. In Fig.~\ref{fig:marginal_param} we show such distributions for three parameters $\{ p^{\bar{d} +\bar{u}}_1, p^{d_v}_2, p^{\bar{u}}_4\}$ together with the corresponding uncertainties from top to bottom: 90\%-symmetric (green), 90\%-asymmetric (blue) and cumulative-$\chi^2$ (red). These parameters were chosen to showcase examples of different behaviour: in the left panel we have a parameter that is Gaussian with the minimal $\chi^2$ sample in the centre of the Gaussian, in the middle panel we see a nearly Gaussian distribution but the minimal $\chi^2$ sample is far away from the centre of the Gaussian, and finally in the right panel the distribution is very non-Gaussian and asymmetric.

In the first case of the $p^{\bar{u}+\bar{d}}_1$ parameter we see that the $\alpha\%$-symmetric and $\alpha\%$-asymmetric estimations agree well. The cumulative $\chi^2$ method shows a much broader uncertainty. For the $p^{d_v}_2$ parameter the $\alpha\%$-symmetric and $\alpha\%$-asymmetric error estimations disagree in their best-fit value and also in their uncertainty estimation, and again the cumulative $\chi^2$ method gives larger uncertainty. Finally, for the $p^{\bar{u}}_4$ parameter the first two methods agree quite well in their uncertainty estimation, but fail to capture the long tail of the distribution. On the other hand, the large span of the cumulative-$\chi^2$ method includes this tail. Generally we can see that the third method shows a different behaviour compared to the other two and is always the most conservative. This can be understood from the method having a different goal: instead of trying to estimate the distribution of the observable of interest, it tries to find the absolute bounds without being in explicit tension with the input experimental measurements. However we can also see that the method lands on the far edges of the parameter distributions and most likely has a significant error attached to its estimate, simply because the absolute sample size in these regions is very low.

\begin{figure}[htb]
	\centering
    \includegraphics[width=0.49\textwidth]{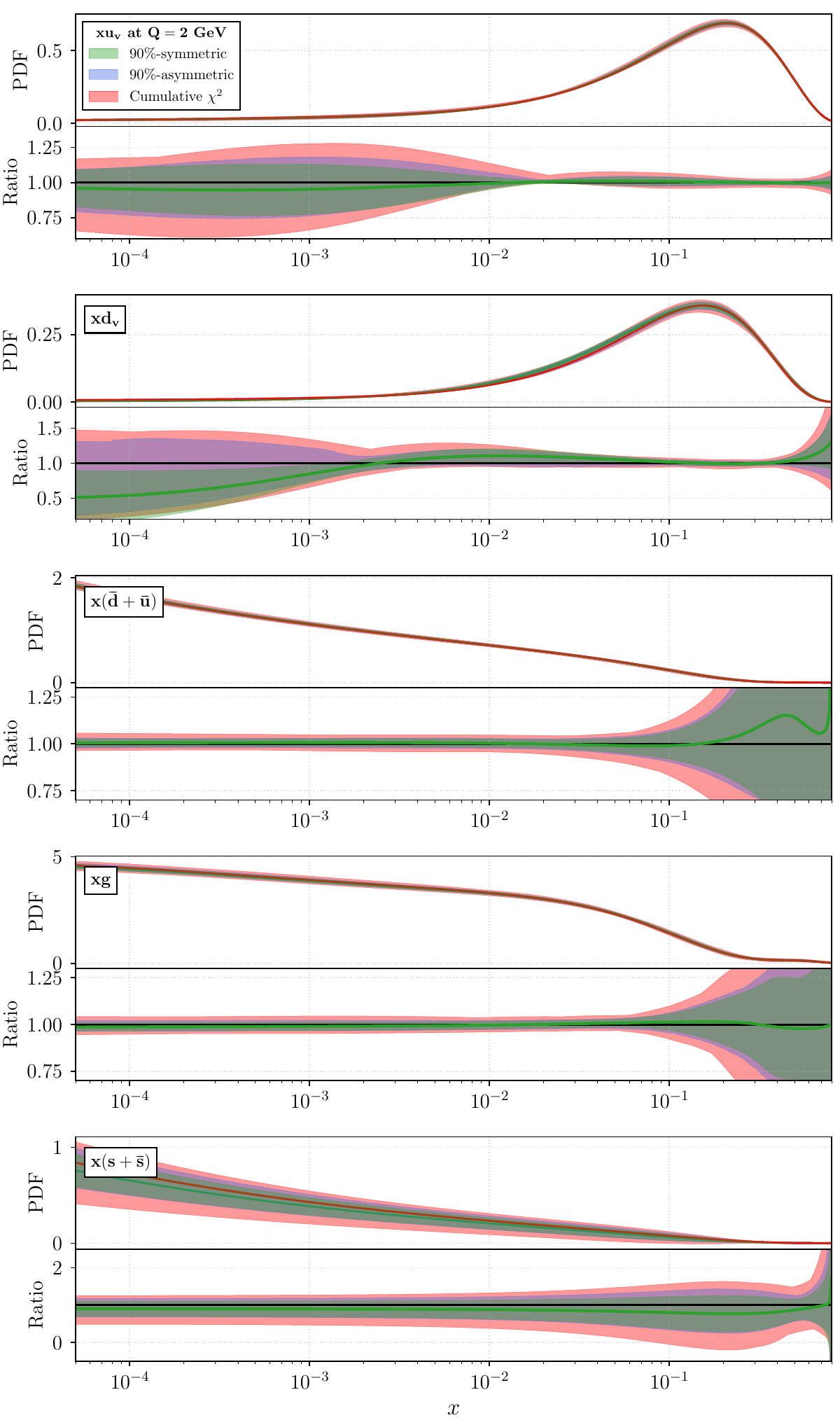}
	\caption{The PDF uncertainties as a function of $x$ for the three methods in the fitted basis for flavour combinations $\{u_v,d_v,\bar{d}+\bar{u},g,s+\bar{s}\}$. The upper plots show the absolute distributions and the lower the ratio to the cumulative $\chi^2$ method.}
	\label{fig:PDF_uncertaintiy}
\end{figure}
In Fig.~\ref{fig:PDF_uncertaintiy} we present uncertainties obtained with the three methods at 90\% confidence level for PDFs. In particular, we plot $u_v$, $d_v$, $\bar{d}+\bar{u}$, $g$, and $s+\bar{s}$ distributions at a scale of $Q=2$ GeV.
On a general level we observe similar behaviour as for individual parameters, \textit{i.e.} we find that the size of the uncertainties increases in the same order: $\alpha\%$-symmetric, $\alpha\%$-asymmetric and cumulative-$\chi^2$ method.
By definition the central values of the $\alpha\%$-asymmetric and the cumulative-$\chi^2$ methods are the same, whereas the central value of the $\alpha\%$-symmetric method can deviate from them. This deviation can be quite large but always is within the error bands of other methods (with the exception of $d_v$ at low-$x$) and the largest deviations happen in $x$ regions where the corresponding distributions are very small.
In fact, the $d_v$ distribution exhibits the largest difference between the methods reaching $\sim 10\%-20\%$. This behaviour is to some extent expected as most of the $d_v$ parameters are non-Gaussian which means that the $\alpha\%$-symmetric method is not appropriate for estimating uncertainties in this case.

\subsection{Comparison with the Hessian method}
\label{sec:MCMC_vs_Hess}
To provide further perspective we compare the results of the MCMC PDF analysis with a traditional approach of a Hessian determination of PDFs, that was briefly reviewed in \cref{subsec:Hess}. We have performed an independent fit to find the minimum of the $\chi^2$ function and then used the Hessian method to estimate PDF uncertainties. The same position of the minimum was used to generate starting points for the MCMC analysis, c.f.~\cref{subsec:startingpts}. Since setting an appropriate value of the tolerance criterion for the Hessian method is non-trivial, and it is one of the fundamental shortcomings of this method, we set it using the findings from our MCMC analysis. This ensures a fair comparison with the MCMC uncertainties, since otherwise the Hessian uncertainties can be arbitrarily scaled by adjusting the tolerance. As the MCMC analysis gave us a reliable $\chi^2$ distribution of the samples, we use it to compute the 90\%-quantile allowing us to reliably define the tolerance, $T^2=21$.
Furthermore, since the computed 90\%-quantile agrees reasonably well with the 90\%-quantile for an exact $\chi^2$-distribution with 15 degrees of freedom, we also expect that the uncertainties resulting from the Hessian analysis will be comparable to those from the cumulative-$\chi^2$ method.

\begin{figure}[htb!]
	\centering
    \includegraphics[width=0.49\textwidth]{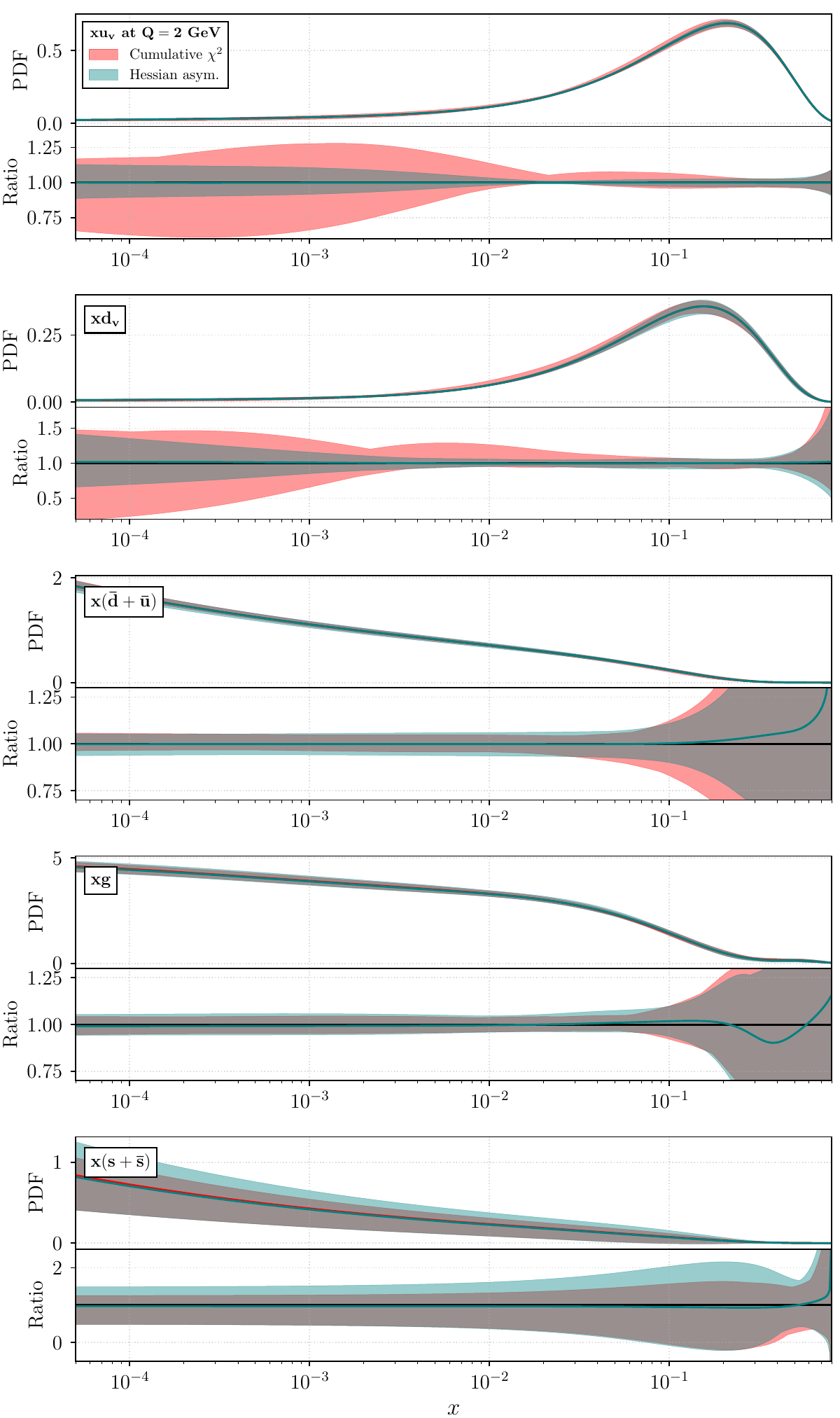}
	\caption{A comparison between the PDF uncertainties from the cumulative $\chi^2$ method versus the Hessian method in the fitted basis as a function of $x$.}
	\label{fig:PDF_MCMC_vs_Hessian}
\end{figure}

In Fig.~\ref{fig:PDF_MCMC_vs_Hessian} we compare PDFs with uncertainties obtained using the cumulative-$\chi^2$ method and obtained from the Hessian fit. We present $u_v$, $d_v$, $\bar{d}+\bar{u}$, $g$, and $s+\bar{s}$ distributions at a scale of $Q=2$ GeV with 90\% confidence level uncertainties.
The first point to note is that the central values obtained with both methods are very close. We observe some deviations but only at large $x$, e.g., for the gluon distribution, but these are small and in the region where the corresponding distributions nearly vanish. This means that the minimum found in the Hessian fit is not far from the MCMC sample with the lowest $\chi^2$, specifically the minimal $\chi^2$ for MCMC samples is $\chi^2_{\mathrm{min}}=2368.25$ ($\chi^2_{\mathrm{min}}/d.o.f=1.20$) and for the Hessian minimum $\chi^2_{\mathrm{min}}=2366.03$ ($\chi^2_{\mathrm{min}}/d.o.f=1.20$).

Much more variation can be observed when inspecting the uncertainties. On the one hand, the MCMC and Hessian uncertainties are quite comparable for gluon and $\bar{d}+\bar{u}$ distributions. This happens especially for $x\lesssim0.1$ where both uncertainties are quite small and the corresponding central values agree very well; for larger $x$ values we see more variation but also the errors are much larger and the variation is always within the uncertainty of both methods. On the other hand, the uncertainties for $u_v$, $d_v$, and $s+\bar{s}$ distributions exhibit rather large differences. Specifically, the uncertainties obtained from the cumulative-$\chi^2$ method are much more asymmetric and in the case of valence distributions larger than the Hessian estimates, whereas for $s+\bar{s}$ they are smaller. 
Particularly, worth noting is the difference between the uncertainties at low-$x$ for $u_v$ and $d_v$ distributions, where the difference can exceed factor 2 and can strongly depend on particular $x$ value.

The behaviour described above is not surprising as most of the marginal distributions of the gluon and $\bar{d}+\bar{u}$ parameters are approximately Gaussian, hence we expect agreement between the two approaches. Conversely, the marginal distributions of the $u_v$, $d_v$, and $s+\bar{s}$ parameters exhibit non-Gaussianity and at the same time large asymmetry. As a result we do not expect symmetric uncertainties on the level of these PDFs. This also means that the Hessian method which assumes Gaussianity can not be reliably applied in such cases. This clearly highlights the advantage of the MCMC estimation of uncertainties which are not limited by Gaussian approximation and can be safely used in more general situations.

\section{Conclusions and Outlook}
\label{sec:conclusions}
We have presented a Markov Chain Monte Carlo extraction of proton PDFs aiming at faithful estimation of PDF uncertainties. We used the adaptive Metropolis-Hastings algorithm to construct 36 Markov Chains which after processing (removing thermalization and autocorrelation) were combined to form a single uncorrelated set of 4068 independent samples/PDFs.
To ensure that the obtained set of samples represents the underlying probability distribution we made sure that each individual chain was sufficiently long to be fully converged and we used the $\Gamma$-method to calculate autocorrelations of individual chains.
We further tested and compared a number of methods for computing uncertainties of PDFs and PDF-dependent quantities based on this set of samples, and concluded that in this case the cumulative-$\chi^2$ method agrees best with the Hessian approach provided that its tolerance is estimated correctly. However, the main result of this study is the chain itself, which gives direct access to the underlying probability distribution of the PDFs and the best method to propagate uncertainties can be chosen based on the observable of interest.

From our analysis it is clear that during the PDF determination we can encounter parameters that are non-Gaussian (see Figs.~\ref{fig:pairwise} and~\ref{fig:marginal_param}). Such situations, especially when the related asymmetry is large (e.g.~parameter $p_4^{u_v}$), invalidates the assumptions of the Hessian method commonly used for extracting PDF uncertainties. 
As a result it challenges the robustness of estimating PDF uncertainties with this method, especially when aiming at percent level precision~\cite{NNPDF:2021njg}.

In order to test the above statement we have performed an independent Hessian analysis using the same setup as for the MCMC analysis (using the same data, same parameters, etc.) and confronted the uncertainties from the Hessian method with those from MCMC. As can be seen from Fig.~\ref{fig:PDF_MCMC_vs_Hessian}, when the assumption of Gaussianity is fulfilled, the uncertainties obtained in both approaches agree well. However, when the PDF parameters describing the distribution of a given flavour deviate from the assumed Gaussian profile (e.g.~for valence distributions) the uncertainty estimates no longer agree. In such cases the Hessian method tends to provide much more symmetric uncertainty bands either underestimating or overestimating the true uncertainties.
On the other hand, the MCMC approach allows for the determination of the underlying probability distribution, and hence provides reliable estimates of the true uncertainties (assuming of course that convergence was obtained and neglecting the parametrisation bias).

Additionally, using MCMC we were able to consistently include a weakly constrained parameter, $p_4^{d_v}$, that would have had to be set to a fixed value in an analysis relying on the Hessian method. 
Further, since MCMC determines the probability distribution around the probed minimum it provides means to reliably estimate the Hessian tolerance based on the quantile of the extracted $\chi^2$ distribution. This can be potentially used to avoid one of the main shortcomings of the Hessian method. 

As stated above, we found that the Hessian method under- or overestimates the PDF uncertainties for non-Gaussian parameters. This is even more relevant for extractions of quantities where the constraints from experimental data are more sparse than for the proton PDF case, since the quadratic expansion usually only holds close to the minimum and less constrains result in a wider available parameter space. One example are nuclear PDFs, where a MCMC study is currently ongoing~\cite{Derakhshanian:2026zkx,Derakhshanian:2023bed,Derakhshanian:2024wxk}. 

Despite the advantages of the MCMC approach in providing a statistically robust treatment of uncertainties, several limitations emerged in our current setup. 
First, in trial setups where we set the posterior to test distributions we noticed that the aMH proposal algorithm fails to explore more than a single local minimum at a time, limiting its ability to probe multimodal posterior landscapes. 
Additionally, we observed poor sampling efficiencies, reflected in low acceptance rates and large autocorrelation lengths, which reduce the effective number of independent samples. 
These issues become more pronounced as the dimensionality of the parameter space increases, highlighting the poor scaling behaviour of our current MCMC implementation. 
An interesting alternative to consider in the future is the Hamiltonian/Hybrid Monte Carlo algorithm~\cite{Duane:1987de} which allows for considerable reduction of autocorrelation but requires many $\chi^2$ evaluations (and its derivative) when proposing a single point.

\section*{Data Availability}

The data that support the findings of this article are openly available \cite{risse_2026_19056773}.

\appendix
%
\section{$\Gamma$-method}
\label{app:GammaMethod}

In this appendix we provide a brief overview of the $\Gamma$-method~\cite{Wolff:2003sm} allowing for the calculation of the ``integrated autocorrelation time'', $\tau$, used in Sec.~\ref{subsec:mcmc}. 

Due to the construction of Markov chains each sample in the chain is strongly correlated with its predecessor. This is called autocorrelation in Monte Carlo time and leads to the fact that the naive formula for the covariance matrix between parameters $x_\alpha$ and $x_\beta$ (i.e. the components $\alpha$ and $\beta$ of $\x$)
\begin{equation}
    C^0_{\alpha\beta} = \big\langle\big(x^i_\alpha-\left\langle x_\alpha\right\rangle\big)\big(x^i_\beta-\left\langle x_\beta\right\rangle\big)\big\rangle
    \label{eq:naive_covariance_matrix}
\end{equation}
underestimates the true values of the full covariance matrix $C_{\alpha\beta}$.\footnote{Here we assume that we only have a single chain. In general the RHS of \cref{eq:naive_covariance_matrix} for $R$ replicas is obtained by the replacement $x_\alpha^{i}\rightarrow x_\alpha^{i,r}$ and $x_\beta^{i}\rightarrow x_\beta^{i,s}$ for the replicas $r$ and $s$. Since the replicas are statistically independent this simply amounts to a Kronecker-$\delta$ on the LHS: $C^0_{\alpha\beta}\rightarrow C^0_{\alpha\beta}\delta_{rs}$, and the analysis can be performed for each replica separately. Only the final result is averaged over the replicas. In our analysis we use the $\Gamma$-method for both the full data set and separately for individual replicas to check for consistency between the replicas.} The latter is defined as
\begin{align}
    C_{\alpha\beta} &= \sum_{t=-\infty}^\infty\Gamma_{\alpha\beta}(t)\notag\\
    &= \Gamma_{\alpha\beta}(0) + 2\sum_{t=1}^\infty\Gamma_{\alpha\beta}(t)
    \label{eq:full_covariance_matrix}
\end{align}
where we defined the autocorrelation function
\begin{equation}
    \Gamma_{\alpha\beta}(t) = \big\langle\big(x^i_\alpha-\left\langle x_\alpha\right\rangle\big)\big(x^{i+t}_\beta-\left\langle x_\beta\right\rangle\big)\big\rangle
\end{equation}
with $\Gamma_{\alpha\beta}(0)=C^0_{\alpha\beta}$. Finally we can define the integrated autocorrelation time $\tau_{{\rm int},\alpha\beta}$ as
\begin{equation}
    \tau_{\rm int,\alpha\beta} = \frac{1}{2}+\sum_{t=1}^\infty\frac{\Gamma_{\alpha\beta}(t)}{\Gamma_{\alpha\beta}(0)}
    \label{eq:autocorrelation_time_definition}
\end{equation}
and reformulate \cref{eq:full_covariance_matrix} to factor out the naive estimate from the autocorrelation
\begin{equation}
    C_{\alpha\beta}=C^0_{\alpha\beta}\,2\tau_{\rm int,\alpha\beta}\,.
\end{equation}
Thus we can see that the naive estimate is scaled by twice the autocorrelation time.\footnote{Note that the squared statistical error on $\langle x_\alpha x_\beta\rangle$ is given by $C_{\alpha\beta}/N$ such that the effective number of samples is roughly reduced to $N/2\tau_{\rm int,\alpha\beta}$ compared to a situation without autocorrelations.}

A direct evaluation of the autocorrelation time from \cref{eq:autocorrelation_time_definition} is not possible because of the infinite sum. The $\Gamma$-method provides a prescription on how to approximate this sum avoiding two main issues.
First, we only have a finite number of samples (the chain has a fixed length) and we therefore have to truncate the sum to some finite window ({\em summation window}) $W$: $\sum_{t=1}^\infty\rightarrow\sum_{t=1}^W$. The second is a signal-to-noise issue: although the correlation between two consecutive samples can be strong, it usually falls off rapidly. Thus at large separation times ($t\gg 1$) we mainly sum over statistical noise and since we can not sum until infinity the noise does not necessarily average to zero. A too small summation window underestimates the signal and a too large summation window overestimates the noise. 
By assuming the autocorrelation to fall off exponentially the $\Gamma$-method defines an algorithm that finds the optimal summation window $W_{\rm opt}$ by explicitly calculating the error contributions to $C_{\alpha\beta}$ from the aforementioned issues, see~\cite[Sec. 3.2]{Wolff:2003sm}. 

The final estimate on the integrated autocorrelation time for the correlation between the parameters $x_\alpha$ and $x_\beta$ is given by
\begin{equation}
    \tau_{\rm int,\alpha\beta} = \frac{1}{2}+\sum_{t=1}^{W_{\rm opt}}\frac{\Gamma_{\alpha\beta}(t)}{\Gamma_{\alpha\beta}(0)}\,.
\end{equation}

In \cref{fig:autocorrelation_vs_thinning}, instead of showing $\tau_{\rm int,\alpha\beta}$ as a function of the summation window $W$ for each parameter combination individually, we use an averaged quantity
\begin{equation}
    \tau_{\rm int}(W) = \frac{1}{2}+\sum_{t=1}^W\frac{\sum_{\alpha\beta}C^0_{\alpha\beta}\Gamma_{\alpha\beta}(t)}{\sum_{\alpha\beta}C^0_{\alpha\beta}}\,,
    \label{eq:tau_int}
\end{equation}
which weighs the autocorrelation times for each parameter combination by its respective correlation strength. This simplification has proven useful in our application. However, we also always checked the autocorrelation times for each individual parameter ($\tau_{\rm int,\alpha\alpha}$) and the possible combinations. The bottom panel of \cref{fig:autocorrelation_vs_thinning} shows the ratio (referred to as the {\em autocorrelation function})
\begin{equation}
    \rho(t) = \frac{\sum_{\alpha\beta}C^0_{\alpha\beta}\Gamma_{\alpha\beta}(t)}{\sum_{\alpha\beta}C^0_{\alpha\beta}}
    \label{eq:rho_W}
\end{equation}
for $t=W$. Since the $\Gamma$-method assumes this function to fall off exponentially, the figure serves as a cross check that this assumption is met.

\begin{acknowledgments}
The authors would like to thank our nCTEQ collaborators, in particular 
Jan Wissmann and Ingo Schienbein
for many useful comments and discussions. 

The work of P.R.~was supported by the U.S.~Department of Energy under Grant \mbox{No.~DE-SC0010129,} 
and by the Office of Science, the Office of Nuclear Physics, within the framework of the Saturated Glue (SURGE) Topical Theory Collaboration.
P.R. thanks the Jefferson Lab for their hospitality. This material is based upon work supported by the U.S. Department of Energy, Office of Science, Office of Nuclear Physics under contract DE-AC05-06OR23177.
The work of P.R., T.J.\ and K.K.\ at the university of M{\"u}nster 
was funded by the DFG through the Research Training Group 2149 ``Strong and Weak Interactions - from Hadrons to Dark Matter''.
A.K. and N.D. acknowledge the support of National Science Centre Poland under the Sonata Bis grant No.~2019/34/E/ST2/00186.
N.D. also acknowledges the support provided by the NAWA-STER Program, project no. PPI/STE/2020/1/00020/U/00001.
\end{acknowledgments}

\bibliographystyle{utphys}
\bibliography{references.bib}
\end{document}